\documentclass[aps,pre,twocolumn,superscriptaddress,floatfix,amsmath]{revtex4-1}
\usepackage[utf8]{inputenc}
\usepackage{psfrag}
\usepackage{epsfig}
\usepackage{graphicx}
\usepackage{color}
\usepackage{amsmath,amssymb}
\usepackage{float}

\begin{document}


\title{Geometrical properties of Potts model during the coarsening regime}

\author{Marcos P. O. Loureiro}
\affiliation{Universit\'e Pierre et Marie Curie -- Paris VI,\\ LPTHE UMR 7589, 4 Place Jussieu, 75252 Paris Cedex 05, France} 

\author{Jeferson J. Arenzon}
\affiliation{Instituto de F\'\i sica and INCT Sistemas Complexos, Universidade Federal do Rio Grande do Sul,\\ CP 15051, 91501-970 Porto Alegre RS, Brazil} 

\author{Leticia F. Cugliandolo}

\affiliation{Universit\'e Pierre et Marie Curie -- Paris VI,\\ LPTHE UMR 7589, 4 Place Jussieu, 75252 Paris Cedex 05, France} 

\date{\today}


\begin{abstract}
We study the dynamic evolution of geometric structures in a poly-degenerate system 
represented by a $q$-state Potts model with non-conserved order parameter 
that is quenched from its disordered into its ordered phase.   
The numerical results obtained with Monte Carlo simulations show a strong relation between the 
statistical properties of hull perimeters in the initial state and during coarsening: the statistics 
and morphology of the 
structures that are larger than the averaged ones are those of the initial state while the ones of small 
structures are determined by the curvature driven dynamic process. We link the hull 
properties to the ones of the areas they enclose. We analyze the  linear von-Neumann--Mullins law, both
for individual domains and on the average, concluding that its validity, for the later case,
is limited to domains with number of sides around 6, while presenting stronger violations in the
former case.  
\end{abstract}

\maketitle


\section{Introduction}
\label{sec:introduction}
 
\begin{figure}
 \includegraphics[width=7cm]{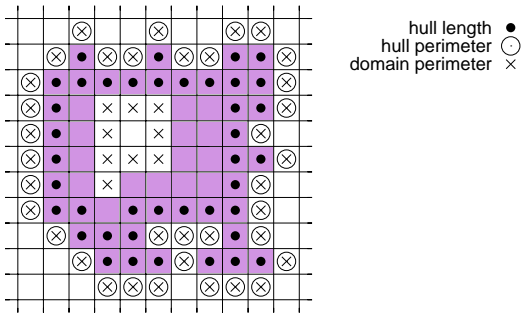}
 \caption{(Color online.) Illustration of various interfacial length definitions on the square lattice. 
 The colored sites are all occupied by the same spin species while spins on white squares are in
 a different state. The hull length (solid circles)  is the external border of the domain. 
 The  hull perimeter (open circles) contains only the external nearest neighbors of the 
 domain (not the internal ones).  The 
 domain perimeter (crosses) is the ensemble of nearest neighbors of the domain and includes the
 internal ones as well. 
 The measurements presented here  are for the hull perimeter only.}
 \label{fig:sketch}
\end{figure}

After a rapid quench across a phase transition,
two or more equilibrium states may compete for the growth of local order;
this reflects in the out of equilibrium evolution 
observed in many different macroscopic systems once they reach the dynamic 
scaling regime~\cite{Bray2002,Puri2004,Cugliandolo2010,Corberi2011}.
In these systems, the properties of macroscopic observables, such as correlation 
functions and linear responses, can still be described in relatively simple terms 
by resorting to the dynamic scaling hypothesis. This 
hypothesis states that averaged macroscopic observables depend upon time only through a 
monotonically growing, time-dependent characteristic length, $R(t)$.
This length, whose growth law depends on the case under consideration, characterizes the 
average linear size of equilibrated patches at each instant $t$. For
example, after being  cooled below their critical temperature magnetic systems form domains 
in which spins are strongly correlated and the magnetization is uniform. 
More precisely, on a lattice, a (geometric) domain is  defined as the ensemble 
of nearest neighbor sites whose spins are aligned. Each of these domains has a
hull or external border, whose length is the number of external sites that are first neighbors of the chosen
domain~\cite{Stauffer1994} and that, in the continuum limit, is referred as the perimeter. 
Figure~\ref{fig:sketch} shows a sketch that illustrates,  on a square lattice,
this and other related definitions.

Systems as diverse as soap froths~\cite{Glazier1990,ThAlGr06,Weaire2009}, cellular tissues and other natural tilings~\cite{MoVaAl90,HoShZi10}, 
magnetic grains or polycrystalline structures~\cite{Babcock1990,FrUd94,Jagla04,Kurtz1980,Kurtz1980a},
type-I superconductors~\cite{PrFiHoCa08}, etc.
are, in a statistical sense, geometrically similar. 
Their overall morphology and growth properties are well described by simple spin models
with multiple ground states. An example is the ferromagnetic Potts model~\cite{Wu1982} whose
variables are spins that take $q$ possible values and interact with their nearest neighbors. The 
coupling exchanges are taken to be isotropic and homogeneous, i.e. independent of the spin variables
and lattice orientation,  
favoring spin alignment at low enough temperature and, therefore, the existence of $q$ degenerate equilibrium 
states below the phase transition. The critical temperature of the bidimensional model on the 
square lattice is known for all values of $q$, $k_{\scriptscriptstyle\rm B}T_c=2J/\ln(1+\sqrt{q})$, but
the nature of the transition depends on $q$, being 
continuous and second-order for $q \leq 4$ while discontinuous for $q>4$. 
When submitted to a sudden quench from the high to the low temperature 
phase, and let subsequently evolve with stochastic non-conserved order parameter (the 
magnetization)  dynamics, the system tends to organize in progressively larger ordered 
structures of each equilibrium state~\cite{Grant1987,Grest1988,OlPeTo04,BeFeCaLoPe07,Loureiro2010}. 
One of the most common applications of this model is to crystal grain growth, where each spin 
configuration represents a different grain orientation and  for $q\to\infty$ the continuum limit is 
realized~\cite{Anderson84,Srolovitz84}.

The simple presence of domains  implies the existence of topological defects, in this case,
the interfaces. When the final quench temperature is not too close to zero 
(to avoid pinning effects when $q>2$) and much lower than the critical temperature (so that the 
starting ordering process does not occur by nucleation when the transition is of 
first order)~\cite{Ferrero2007,Ferrero09}, at 
sufficiently long times the walls around large domains tend to be flat and  
the evolution (with non-conserved order parameter) is driven by their curvature.
Indeed, the surface tension implies a force per unit area that is proportional to the local 
mean curvature, $\kappa$, acting on each point on the wall. This force is responsible 
for the motion of the interfaces. The field theoretical description (\`a la time-dependent Ginzburg-
Landau equation) in the continuum leads to the Allen-Cahn equation for the 
local velocity, 
${\vec v}=-(\lambda_q/2\pi)\kappa {\vec n_s}$, where $\lambda_q$ is a temperature and 
$q$-dependent dimensional constant related to the surface tension and mobility of a domain wall 
and $\vec n_s$ is a unit vector normal to the 
surface~\cite{AllenCahn}. 
The sign indicates that the velocity points in the direction that tends to reduce the local 
curvature. Thermal effects play a minor role, affecting only the constant $\lambda_q$. 
The time-dependence of the area contained within any finite interface can be deduced by 
integrating the velocity around the hull, i.e. the external interface. This calculation is specially 
simple in $d=2$ since one can 
use the Gauss-Bonnet theorem to find a close expression for the 
area rate of change, that can be written in the following 
compact form:
\begin{equation}
 \frac{dA}{dt} = \oint {\vec v}  \wedge d{\vec l} = 
                                                    \frac{\lambda_q}{5} \left( n-6 \right)
                                                    \;\; \mbox{if} \;\;\;q\geq 2 \; ,                                             
\label{eq:dAdt}
\end{equation}
where $n$ is the number of sides of the external interface, each side being the border between the chosen
area $A$ and one of its neighboring domains. We redefined 
$\lambda_q$ for convenience. 
While all non-percolating hull-enclosed areas of a system with $q=2$ states have only one side, for $q>2$ 
hull-enclosed areas with more than one side can exist and $n$ can be larger than one. In the latter case, the 
equation yields the von Neumann-Mullins law for the area $A$ of an $n$-sided hull-enclosed area~\cite{vonNeumann1952,Mullins56}. 
As a result, this area can grow, shrink or remain constant depending upon the number of 
sides being larger than, smaller than or equal to 6, respectively. Moreover, in the course of 
the evolution, the number of  sides that a given external interface has can change,  so that the 
equation ruling the area evolution changes through the time dependence in $n$, a function that 
one cannot characterize in full detail. It is also clear that for $q>2$  areas do not evolve
independently of each other, as occurs for $q=2$, as $n$
is a quantity that also affects the behavior of the neighboring areas.

For $q=2$, i.e.~the bi-dimensional Ising model, the time-dependence of a given
hull-enclosed area 
can be easily determined from the integration of Eq.~(\ref{eq:dAdt}) since $n(t)=1$ for all 
areas and these evolve independently. The number of hulls with enclosed area $A$ 
per unit system area ($A \gg A_0$, where $A_0$ is a small area cut-off) in the interval $(A,A+dA)$ at time $t$ is related to 
the distribution 
at the initial time $t_0$ through, $n_h(A,t)=n_h \left[ A + \lambda_h (t-t_0), t_0 \right]$. The two 
natural choices for the initial states are equilibrium at $T_0=T_c$ or $T_0\to\infty$ (all high temperature 
initial conditions become equivalent to the latter). In the former 
case the distribution $n_h$ is known analytically and in the latter it is very close to the one 
of critical percolation that is also known exactly~\cite{CardyZiff2003}. Therefore
the dynamic distributions are known as well~\cite{Arenzon2007,Sicilia2007}.  More precisely, 
the equilibrium
critical Ising hull-enclosed area distribution is 
$n_h (A,0)=c_h/A^2$ with $c_h = 1/(8\pi \sqrt{3})$ and one finds
\begin{equation}
 n_h (A,t)= \frac{c_h}{[A+R^2(t)]^2}
 \; , 
 \label{eq:criticalIsing}
\end{equation}
for $t\gg t_0$. 
Note that this expression exhibits the scaling form, $n_h (A,t) = t^{-2}f(A/t)$
since the characteristic length scales as $t^{1/2}$. For a quench from infinite temperature
one observes  that  the distribution takes the critical form of random 
percolation in a few time steps~\cite{Arenzon2007,Sicilia2007}, 
even though the initial fraction of $+1$ (or, equivalently, $-1$) spins, 50\%,  
is smaller than the critical density of random percolation on a square lattice, $\rho_c \sim 0.59$. 
This feature was also exploited in~\cite{Barros2009} to explain the existence of percolating blocked 
asymptotic states in the zero-temperature dynamics of this model.

Unlike in the $q=2$ case, $n_h(A,t)$ is not analytically known for $q>2$. To start with, there
is no simple relation between the area distribution at time $t$ and the one at the initial time $t_0$ 
and, moreover, we expect the distribution to get scrambled in a non-trivial way during the 
coarsening process. Thus, in a previous work~\cite{Loureiro2010} we studied this problem 
with Monte Carlo simulations. 
First, we confirmed that the coarsening process is characterized by a 
growing length that depends on time as~\cite{Grant1987,Grest1988} 
\begin{equation}
R^2(t) \simeq \lambda_q \ t
\label{eq:growing-length}
\end{equation}
by studying different correlation functions and their scaling properties. Second, we investigated
the hull-enclosed area distribution.  We found that in the cases in which the transition is second 
order ($2\leq q\leq 4$) and the initial configuration is critical, $T_0=T_c$, the
dynamic distribution has a power law tail for sufficiently large areas.
Quantitatively, we found that the exponent is $2$ independently of $q$ (within our numerical accuracy)
while the prefactor depends on $q$. Interestingly enough, even though the individual 
area rate of change depends on the number of sides, the long-range correlations present in 
the initial state are preserved and the system keeps the distribution shape during evolution. 
Assuming that the number of sides in the von Neumann-Mullins equation~(\ref{eq:dAdt}) can be 
replaced by a constant mean, $n(t) \rightarrow \langle n \rangle$, and using 
Eq.~(\ref{eq:criticalIsing}) at $t=t_0$ we proposed
\begin{equation}
 n_h(A,t) \simeq \frac{(q-1)c^{(q)}_h}{[A+R^2(t)]^2}
 \;,
 \label{eq:criticalPotts}
\end{equation}
with $c^{(q)}_h$ $q$-dependent constants. Within our numerical accuracy, this form describes well
the hull-enclosed area distribution.
The results are very different in the case of infinite temperature initial conditions. An uncorrelated 
spin configuration representative of equilibrium at $T_0\to\infty$ can be mapped onto one of the
random percolation model. On the square lattice the species density, $\rho=1/q$, 
is much smaller than the critical random percolation limit and the distribution does not become 
critical nor acquire a power-law decay at any time. Instead, it has an exponential tail. 
A similar behavior 
is observed for $q>4$ even when the initial state is taken from equilibrium at $T_c$. The short-range 
correlations initially present become irrelevant after a finite time and the system loses memory of 
the initial state. See also the results in~\cite{Lukierska2010}.

In the nearest-neighbour Potts model considered here, diversely from the
so called cellular Potts model in which several layers of interacting neighbours
are considered, the lattice anisotropy is an important ingredient. This fact,
together with the small number of colors (thus coalescence may be important)
deviate the model considered here from the ideal grain growth situation. 
Under these conditions, it is interesting to see whether the areas and
perimeter distributions have any resemblance with those obtained with
different methods or experimentally. Moreover, the von Neumann-Mullins law, when
applied to individual grains, also depends on the ideal grain growth 
hypothesis.
Thus, in this paper we investigate additional geometric properties of the non-conserved order parameter 
dynamics of the ferromagnetic Potts model through Monte Carlo simulations on a square lattice 
with linear size $L=1000$. The statistical analysis was
performed using over 1000 samples, enough to reduce the errorbars to values that are
smaller than the data points. Each case was studied by considering an initial equilibrium 
state at $T_0=T_c$ or an uncorrelated one ($T_0 \rightarrow \infty $).
The critical state ($T_0=T_c$ and $2\leq q < 4$) 
was obtained by running 500 to 4000 Swendsen-Wang algorithm steps, while the 
uncorrelated one was created by randomly choosing the state of each of the 
$N=L^2$ spins on the lattice. After reaching equilibrium, the system was suddenly quenched to
$T_f=T_c/2$ where we expect pinning and nucleation effects not to be present~\cite{Ferrero2007,Ferrero09} and  
the subsequent evolution to be a curvature-driven process. Time is measured in Monte Carlo units (MCs), that
is $L^2$ single spin flips such that, statistically, each spin is updated once in every step. 

The paper is structured as follows. In Sec.~\ref{sec:discussion} we present a qualitative discussion
of the problem with emphasis upon the role played by the initial conditions.  
In Sec.~\ref{sec:relations} we discuss the relation between hull-enclosed areas and 
perimeters and their rates of changes. 
We underline the relation between the domain morphology and the characteristic length $R(t)$
according to the temperature of the initial state. In Sect.~\ref{sec:rateofchange} we put the von Neumann-Mullins equation to the test.
In Sec.~\ref{sec:distribution} we show our numerical
results for the time evolution of the hull perimeter distributions.
Finally in Sec.~\ref{sec:conclusions} we summarize the results and we conclude.


\section{Initial conditions}
\label{sec:discussion}

\begin{figure}[t!]

 \centerline{$T_0\to\infty$ \hspace{3cm} $T_0=T_c$}

 \includegraphics[width=40mm]{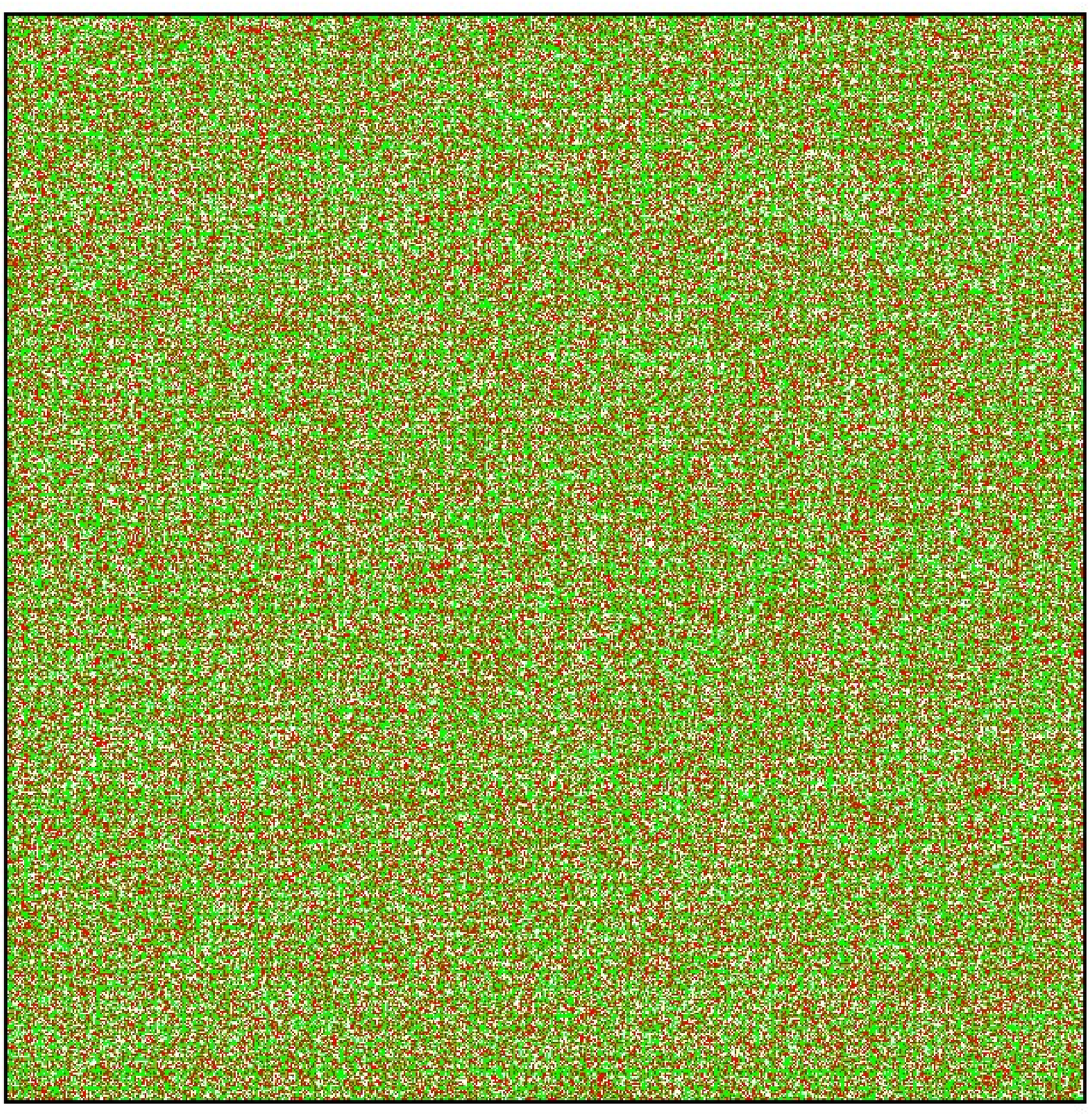}
 \includegraphics[width=40mm]{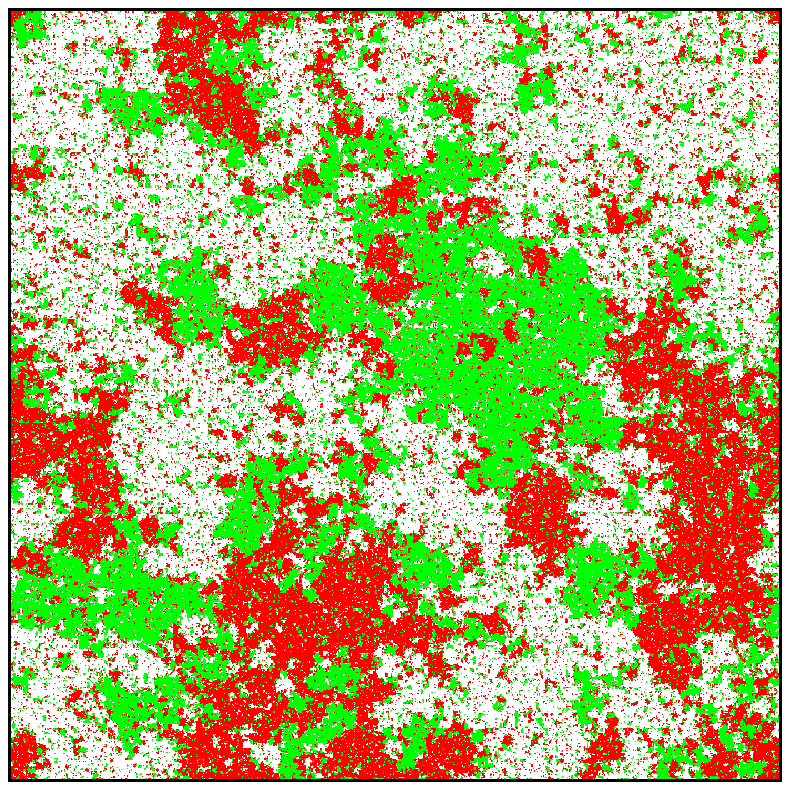}

 \centerline{(a)\hspace{4cm}(b)}

 \includegraphics[width=40mm]{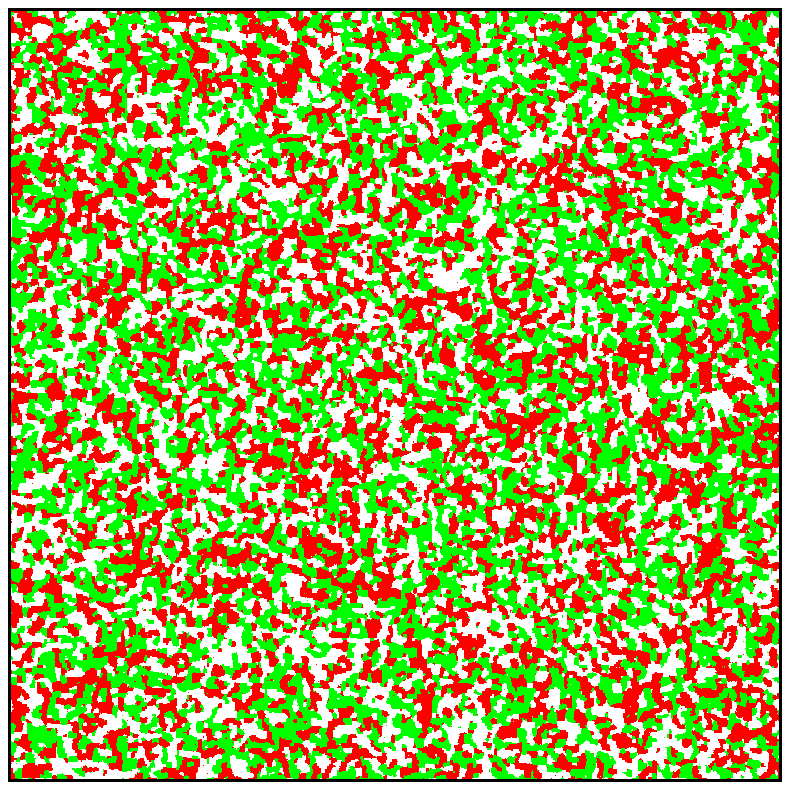}
 \includegraphics[width=40mm]{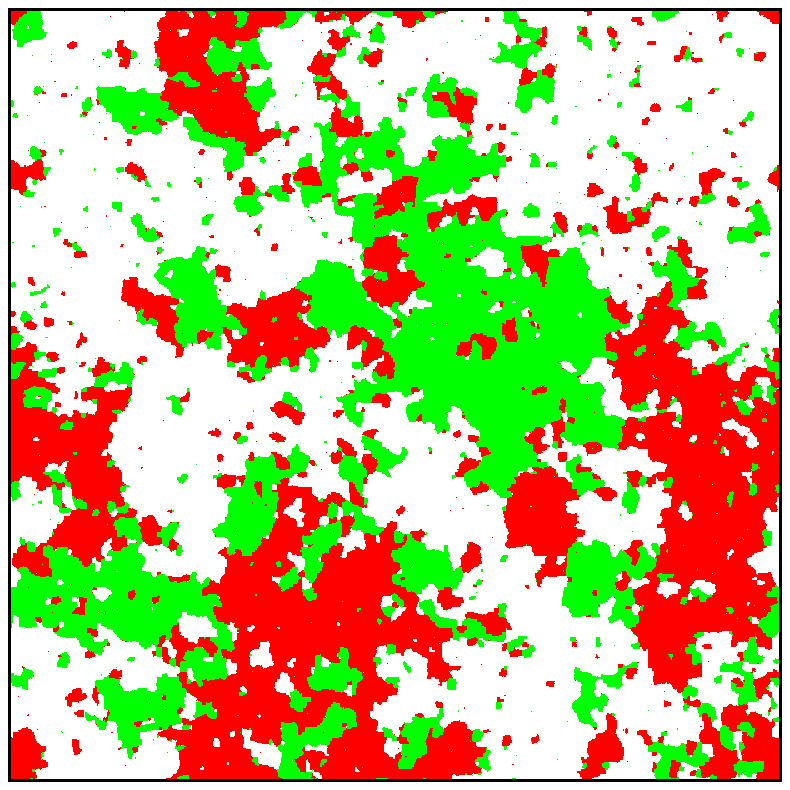}

 \centerline{(c)\hspace{4cm}(d)}
 
 \includegraphics[width=40mm]{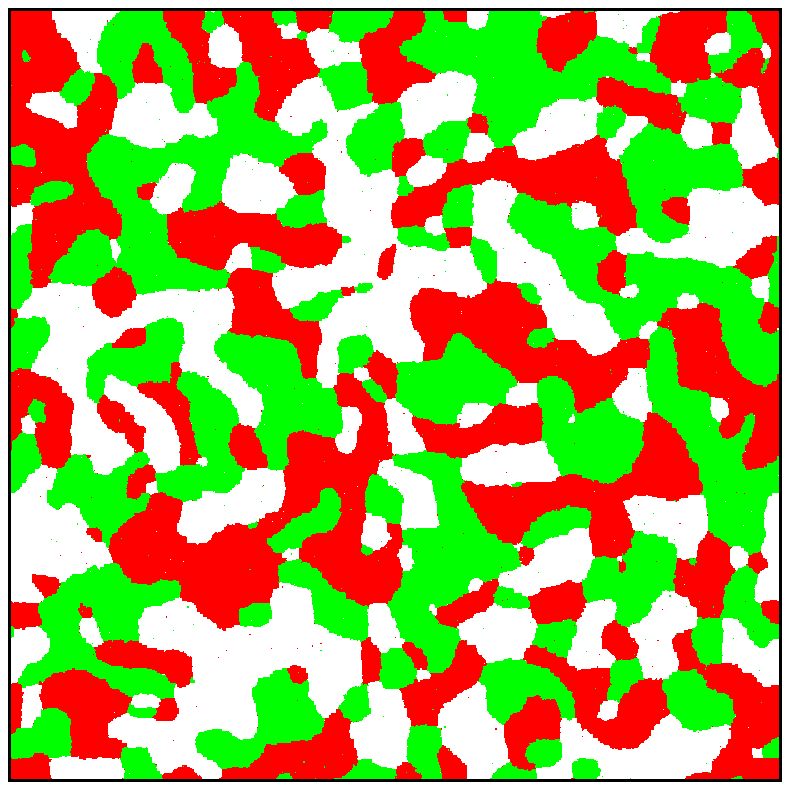}
 \includegraphics[width=40mm]{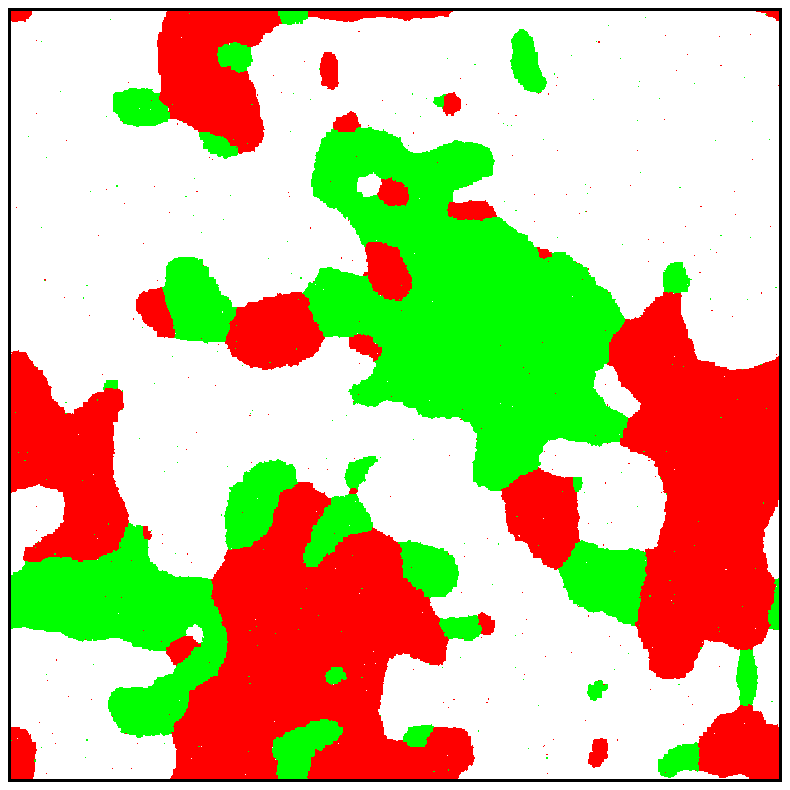}

 \centerline{(e)\hspace{4cm}(f)}
 
 \includegraphics[width=40mm]{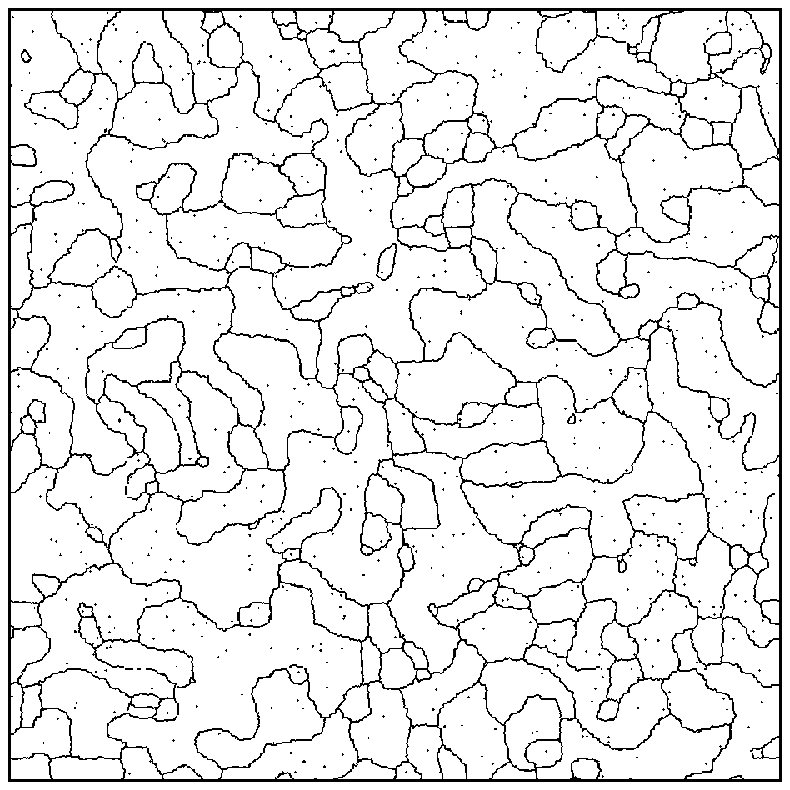}
 \includegraphics[width=40mm]{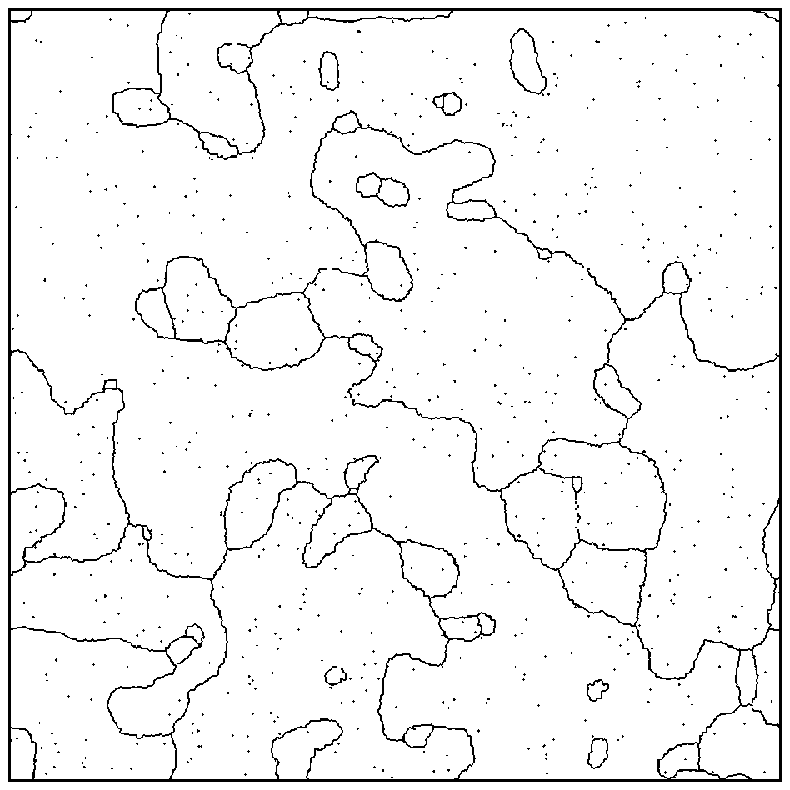}

 \centerline{(g)\hspace{4cm}(h)}
 
 \caption{(Color online.) Snapshots of the $q=3$ ferromagnetic Potts model on a square lattice 
 with linear size $L=1000$ after a quench, from initial states prepared at $T_0 \rightarrow \infty$
(left column) and $T_0=T_c$ (right column), to the final temperature  $T_f=T_c/2$. Times shown
are, from top to bottom, $t=0, \ 2^5$ and $2^{10}$ MCs,  respectively. 
In panels (g) and (h) the interfaces in the configurations at the 
 latest time $t=2^{10}$ MCs [(e) and (f)] are shown, along with some small thermal fluctuations.}
\label{fig:snapshotq3}
\end{figure}

\begin{figure}

 \centerline{$T_0\to\infty$ \hspace{3cm} $T_0=T_c$}

 \includegraphics[width=40mm]{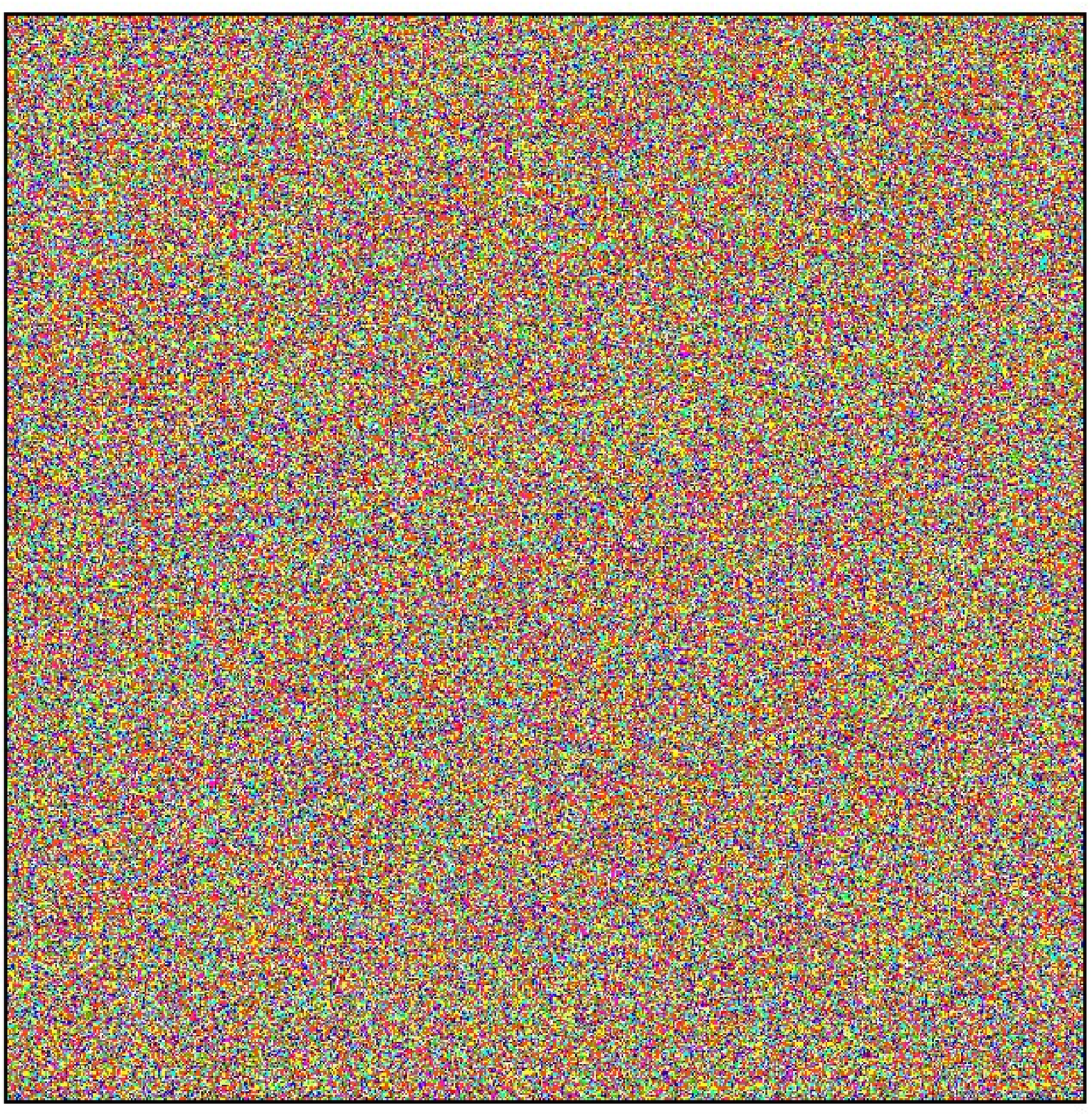}
 \includegraphics[width=40mm]{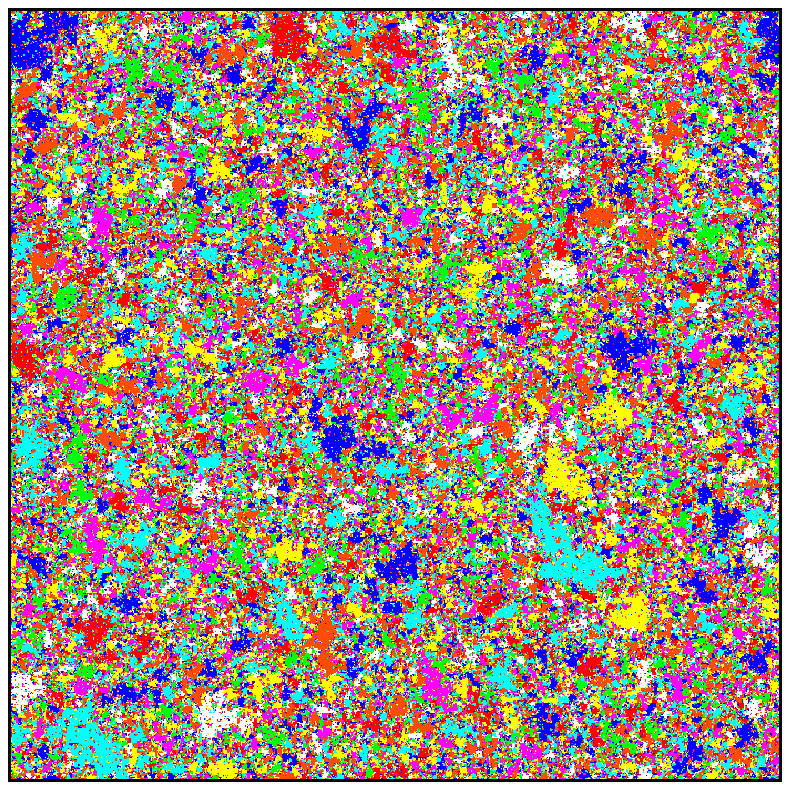}

 \centerline{(a)\hspace{4cm}(b)}

 \includegraphics[width=40mm]{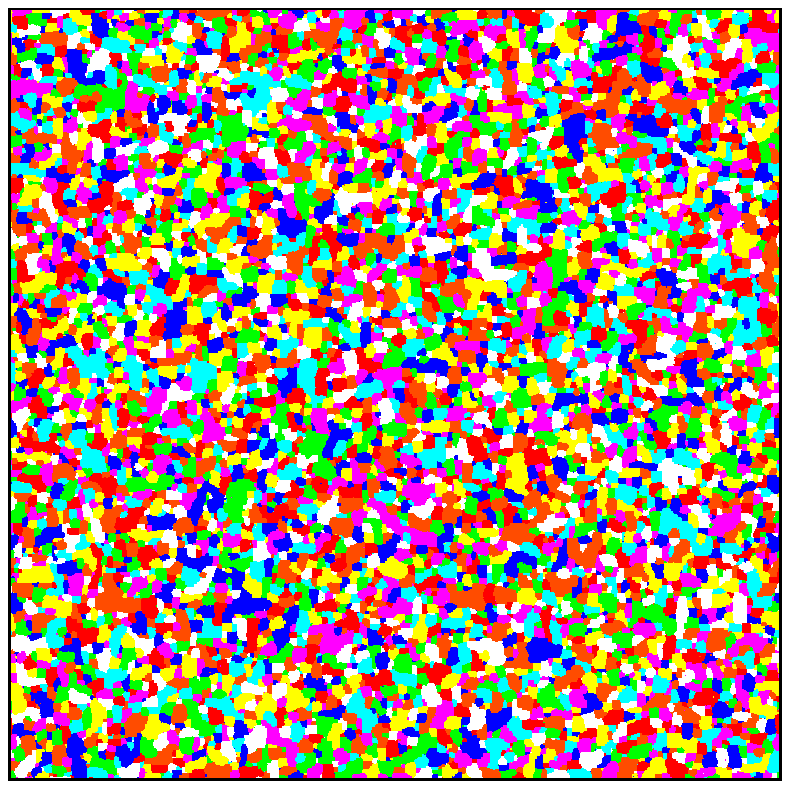}
 \includegraphics[width=40mm]{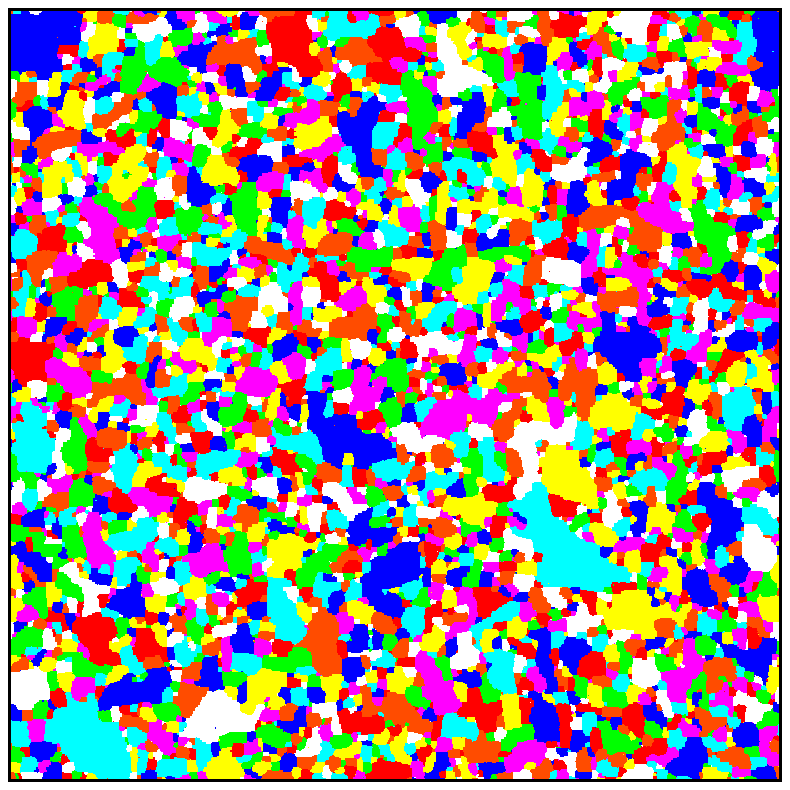}

 \centerline{(c)\hspace{4cm}(d)}
 
 \includegraphics[width=40mm]{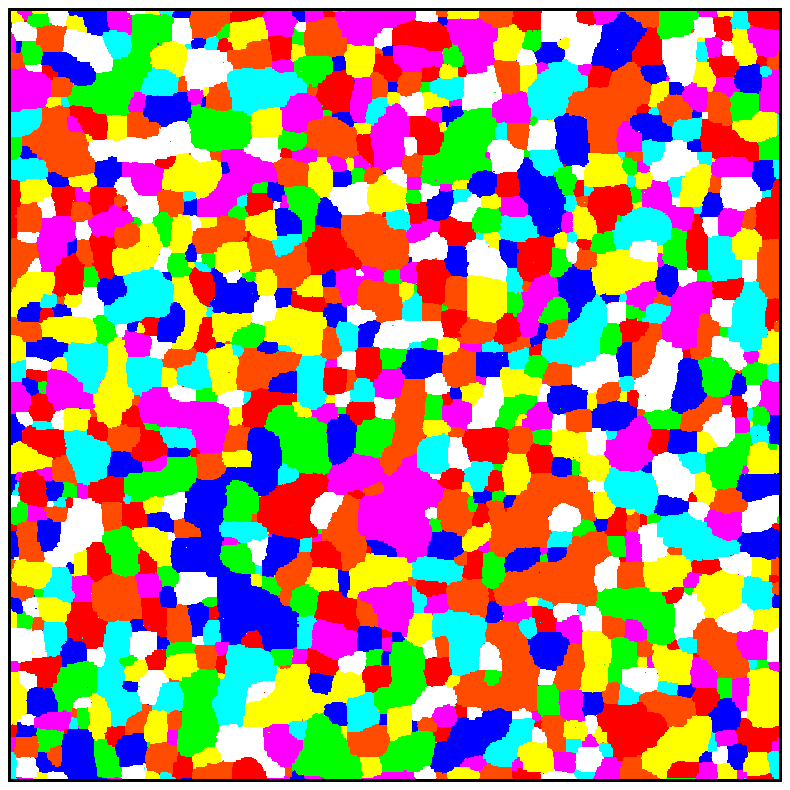}
 \includegraphics[width=40mm]{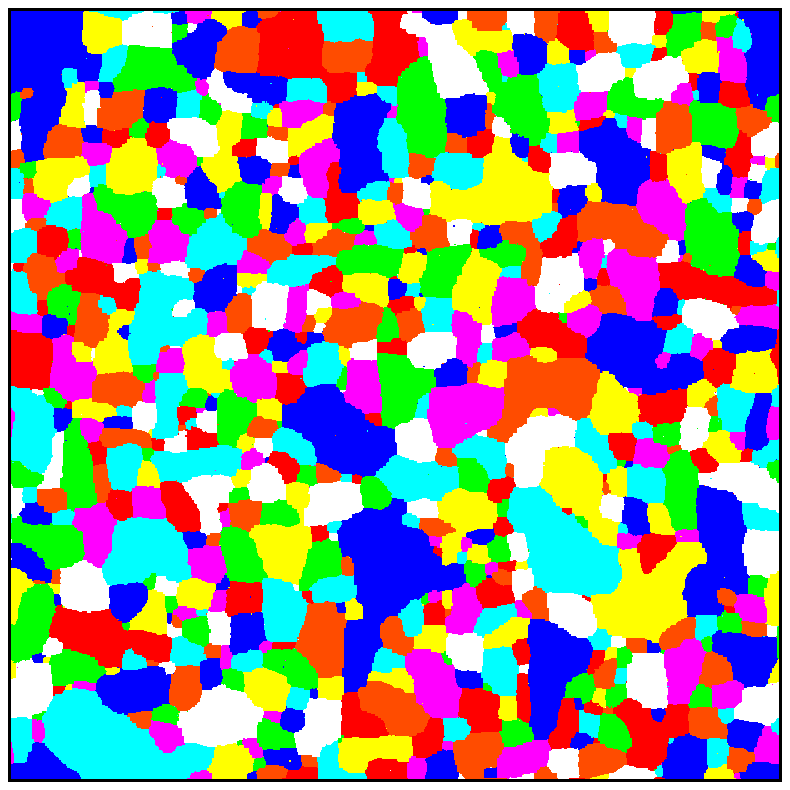}

 \centerline{(e)\hspace{4cm}(f)}
 
 \includegraphics[width=40mm]{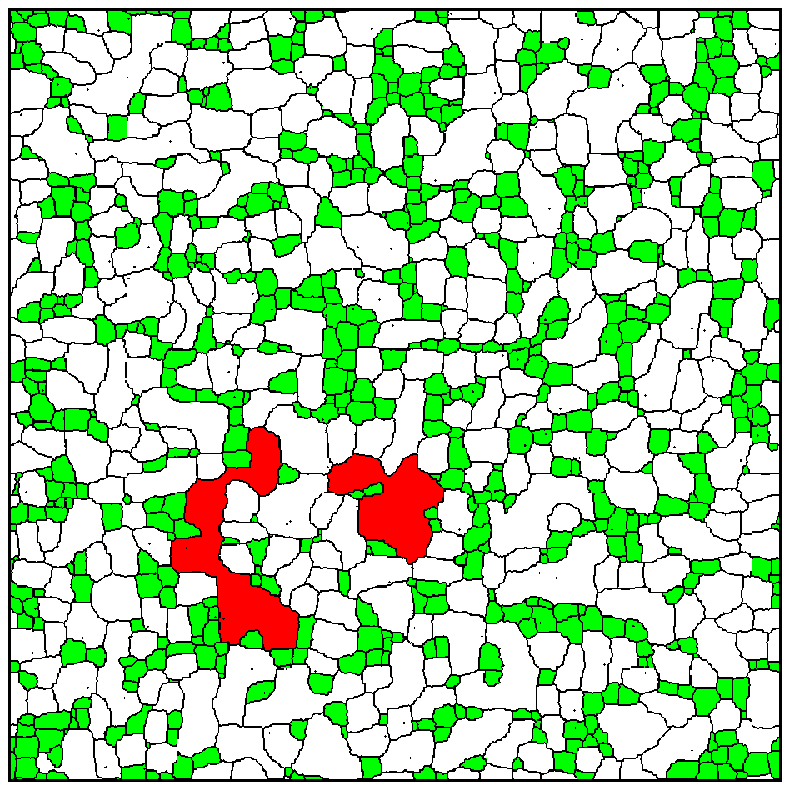}
 \includegraphics[width=40mm]{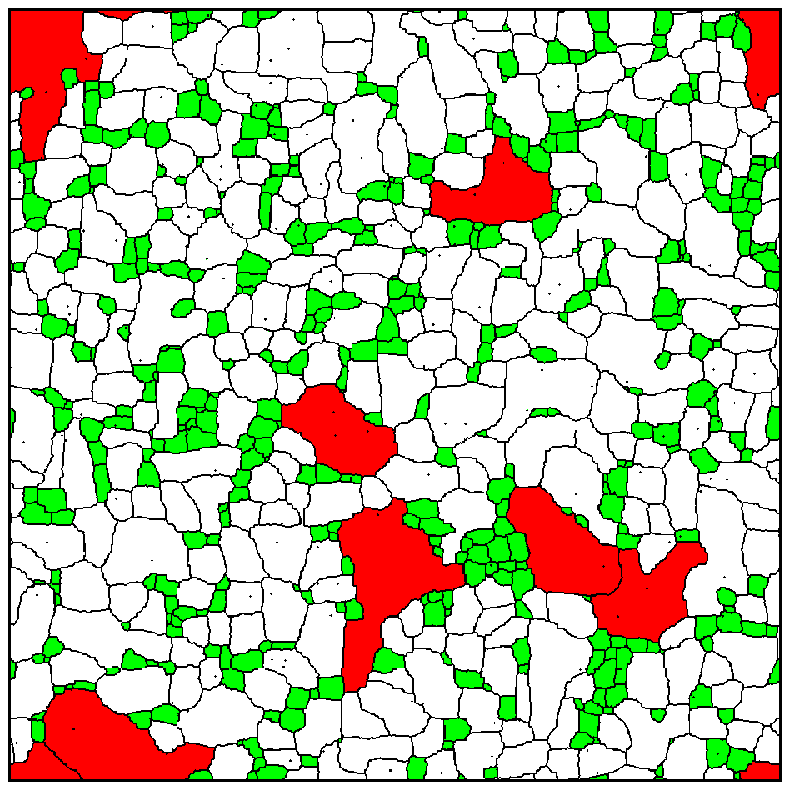}
 
 \centerline{(g)\hspace{4cm}(h)}
 
 \caption{(Color online.) Sequence of snapshots for the $q=8$ Potts model. The left column
 displays results obtained by using an equilibrium initial configuration at $T_0\to\infty$ (a) 
 and its time evolved after the quench [(c) and (e)]. 
 The right column shows configurations found by using an equilibrium initial state at 
 $T_0=T_c$ (b) and after the quench [(d) and 
 (f)]. From top to bottom the measuring times are $t=0, \ 2^7$ and $2^{10}$ MCs, respectively, and 
 $T_f=T_c/2$. In panels (g) and (h) we show the domains with areas $A<R^2$ (green, or light grey), 
 $R^2(t) < A < 10 R^2(t)$ (white)
 and $A>10 \ R^2(t)$ (red, or dark grey) for the configurations at $t=1024$ MCs.}
\label{fig:snapshotq8}
\end{figure}

The initial conditions, either at critical ($2\leq q\leq 4$) or infinite temperature ($\forall \ q$), have
very different characteristics and thus play an important role in 
the system's evolution after the temperature quench, leading
to very different dynamic structures. 
When a $2\leq q \leq 4$ model is in equilibrium 
at $T_c$ its configuration has long-range correlations. In 
$d=2$ the thermodynamic transition also corresponds to a percolation 
transition and one spanning cluster is already present in such initial conditions~\cite{Vanderzande1989,Saleur1987,Janke2003,Picco2009}.
Naturally, such spanning clusters are not counted in our analysis since their perimeters would 
be severely under-estimated due to the finite system size. On the other hand, when the system presents a first order phase transition ($q>4$), 
the correlations are short-ranged at $T_0=T_c$ and the spanning cluster is not present initially nor at any time within
the timespan (and system size) considered in our study.
When the  initial state is one of equilibrium at infinite temperature,
$T_0 \rightarrow \infty$, the correlations are absent and 
the model can be mapped onto  random percolation. 
No spanning cluster is present initially for $q>2$ and none is generated during evolution 
(the case $q=2$ is subtly different, see~\cite{Arenzon2007,Sicilia2007,Barros2009}).
These distinctive features are made evident in the 
snapshots shown in Figs.~\ref{fig:snapshotq3} ($q=3$) and~\ref{fig:snapshotq8} ($q=8$), 
for the two initial conditions, infinite (left column) and critical 
temperature (right column).

For $q=3$, Fig.~\ref{fig:snapshotq3}, the absence of correlations at $T_0\to\infty$ is clear from 
panel~(a) while the long-range correlations and a spanning cluster present at $T_c$ can be
easily visualized in panel~(b).
In panels (c) and (e) the snapshots show the thermal evolution after a quench from 
$T_0\to\infty$ to $T_f=T_c/2$ at times $t=2^5$ and $2^{10}$ MCs. In (d) and (f) the snapshots show the 
configurations 
at  the same times after a quench from $T_0=T_c$. The perimeters of the domains at $t=2^{10}$
MCs are shown in (g) and (h). The right column panels demonstrate that the seeds for the large 
dynamic domains were already present in the initial condition and, although these objects 
do not translate during the evolution, they get rid of small, internal domains.
Indeed, in the explored timespan, basically the modifications are that small domains are erased and the walls get
smoother. A similar conclusion cannot be drawn from the left 
column snapshots as it is difficult to identify a seed for the structures seen at the latest time 
from the very disordered initial condition. 
A common feature of both $T_0=T_c$ and $T_0\to\infty$ evolved configurations is that  
domains with small areas have an approximately   round shape.
Instead,  for large areas the pattern  is rougher and one finds that 
these are sometimes stretched.
A crossover in the morphology of the domains is determined by the 
characteristic length $R(t)$ and it will be
discussed in Sec.~\ref{sec:relations}.


Figure~\ref{fig:snapshotq8} shows a sequence of snapshots of the $q=8$ Potts model 
with both $T_0\to\infty$ (left column) and $T_0=T_c$ (right column) initial conditions. The quench 
is done to $T_f=T_c/2$, and the configurations at
different times, $t=128$ MCs (second row) and $t=1024$ MCs (third row) are shown.
Since none of the initial configurations is critical, thus having either
a vanishing ($T_0\to\infty$) or a finite ($T_0=T_c$) correlation length, one might 
expect them to be statistically equivalent.
However, the dynamically evolved configurations are different
in at least two senses. First, the domains evolved from $T_c$ are larger, as can be 
clearly seen at naked eye. More importantly, the shape of the large domains are very different. 
Panels (g) and (h) take the configurations in panels (e) and (f) and color in the same way 
domains with areas $A<cR^2$ (green), $cR^2 < A < 10R^2$ (white)
and $A>10R^2$ (red). In this way we highlight the density of domains with small, intermediate 
and large size as compared to the typical one, $R^2$, at the measuring time.
On the one hand, when the initial state is uncorrelated (left column panels) few domains with area
$A>10R^2$ survive at the latest time (two) and those have a stretched form. 
On the other hand, when the initial state is critical
(panels in the right column) more domains remain at the chosen time (five), they have larger 
area and are usually less stretched. The difference will be made quantitative in 
Sec.~\ref{subsec:scatter}. It is clear from these figures that it is very hard
to collect good statistics for large areas. 


\section{Areas and Perimeters}
 \label{sec:relations}
 
 In this section we analyze the relation between hull-enclosed areas and 
hull perimeters. 
For any sensible definition of volume and interface,
the volume $V$ of a compact domain with a compact interface $S$ should satisfy
\begin{equation}
V \sim p^d 
\; , 
\qquad\qquad S \sim p^{d-1} 
\; , 
\label{eq:Ap2}
\end{equation}
with $p$ a linear dimension and 
$d$ the dimension of space, leading to 
\begin{equation}
S \sim V^{1-1/d}
\; .
\end{equation}
However, 
if a domain has a fractal surface or holes (other domains) in its interior, 
Eqs.~(\ref{eq:Ap2}) are not necessarily valid~\cite{Stauffer1994}. In practice, 
the presence of holes inside domains depends on the dimension of space and the number of 
ground states of the system. While for $d=2$ and $q=2$ many domains have other
internal domains~\cite{Sicilia2007}, 
for $d=2$ and $q\geq3$ internal domains are much rarer (this can be checked from 
the snapshots in Fig.~\ref{fig:snapshotq8} for $q=8$). Therefore, domain 
areas and hull-enclosed areas differ by very little for $q\geq 3$~\cite{Loureiro2010}.
In the rest of this paper we focus on hull-enclosed areas and in this Section on their relation 
to the hull perimeters themselves.

\begin{figure}
 \includegraphics[width=25mm]{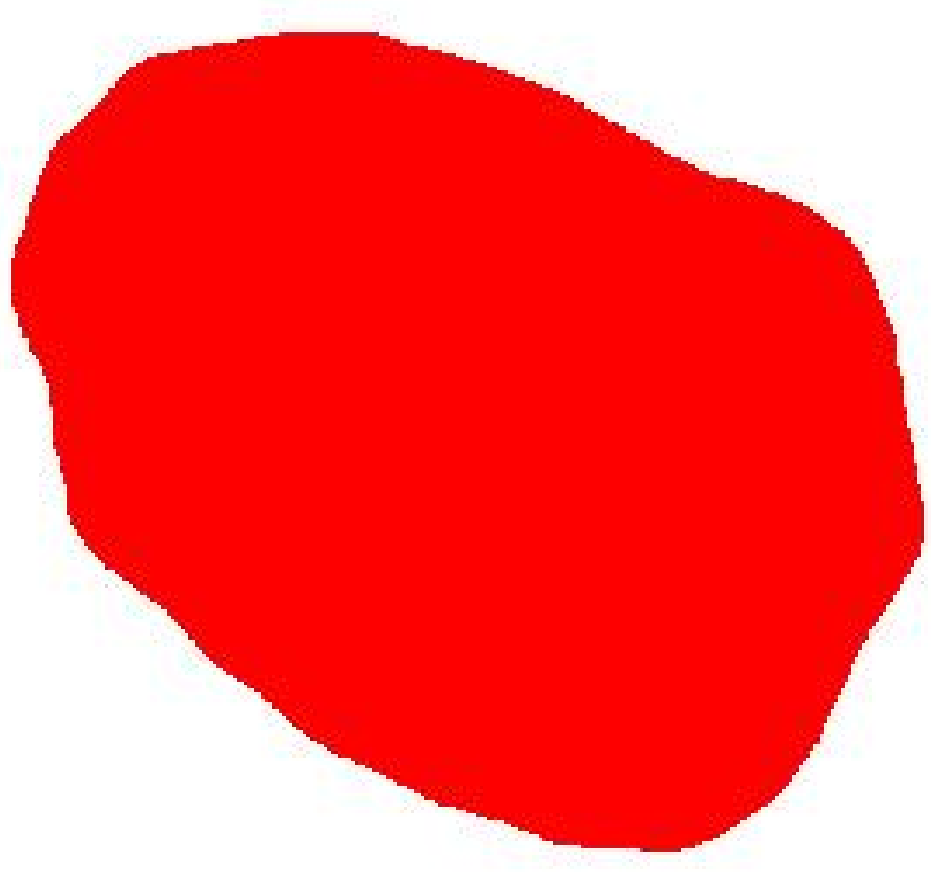}
 \includegraphics[width=25mm]{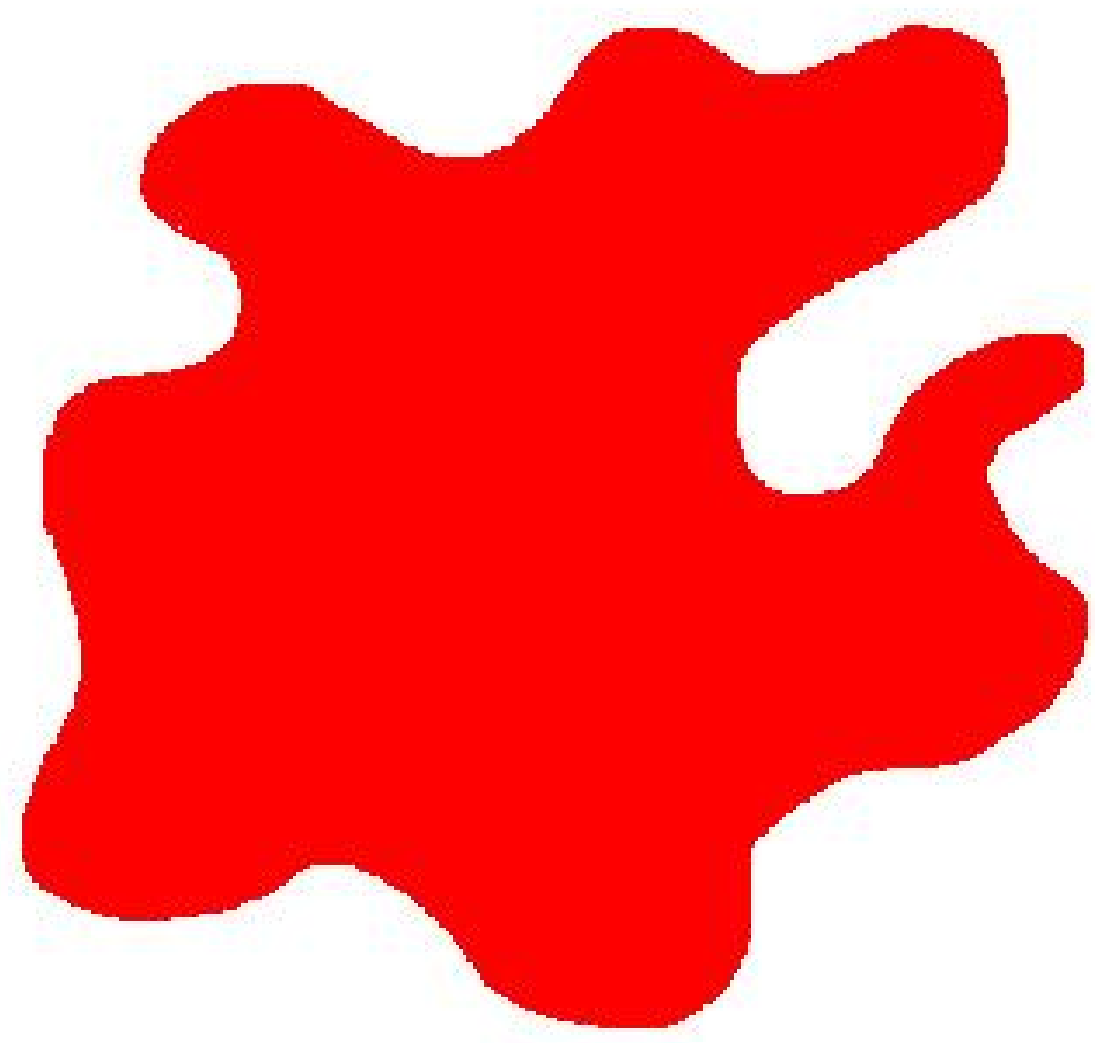}
 \includegraphics[width=25mm]{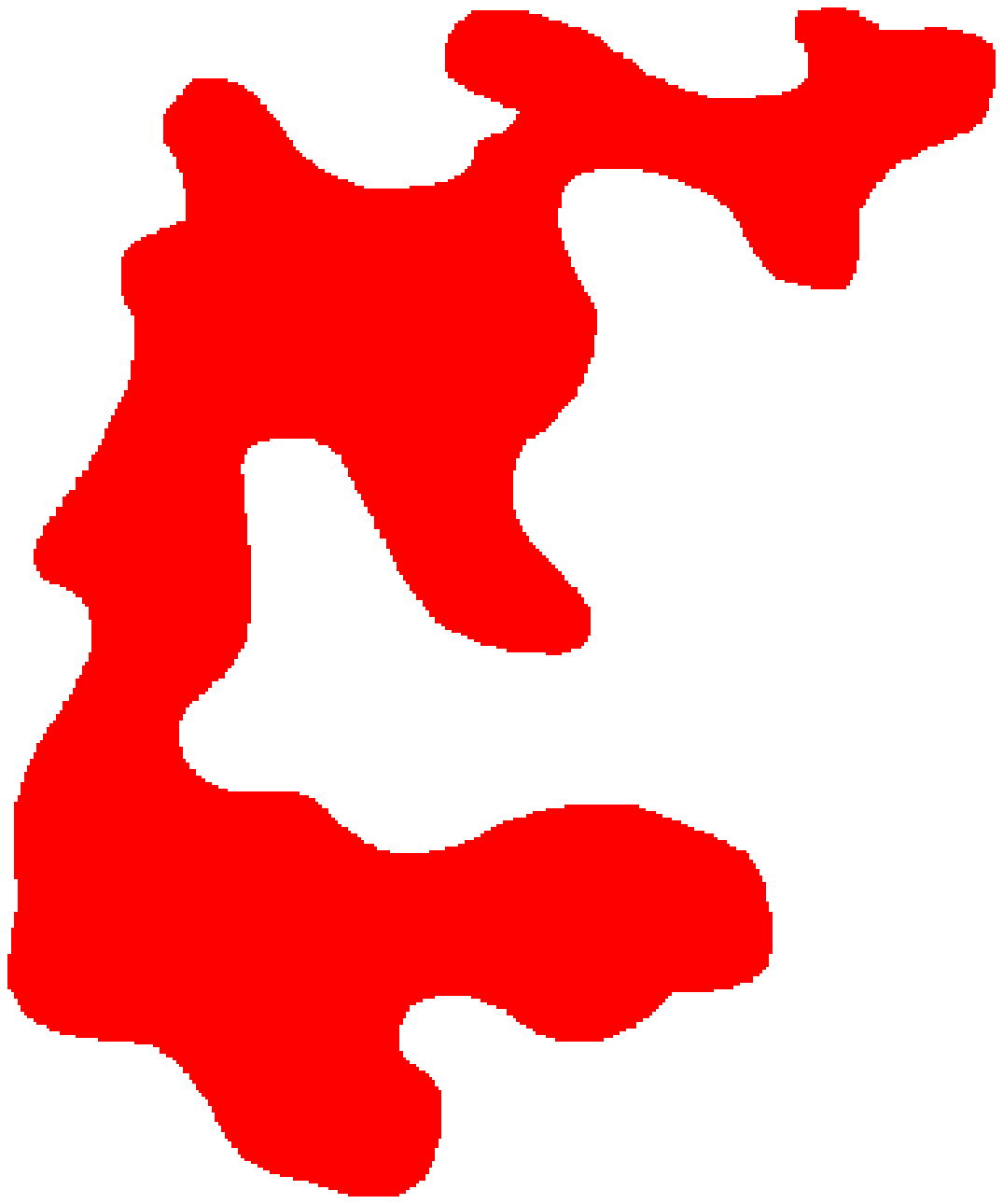} \\
 \centerline{(a)\hspace{20mm}(b)\hspace{20mm}(c)}
 \caption{(Color online.) 
 Examples of typical domain structures obtained in the simulations. In (a) a regular domain with area 
 $A\sim p^{\alpha_h}$ and $\alpha_h=2$. The domain in~(b) has a rougher morphology  
 and a smaller  exponent ${\alpha_h}$ certifies this feature, 
$1<{\alpha_h}<2$. In (c) a domain with very rough and stretched morphology, $A \sim p$; the exponent
${\alpha_h}$ reaches it minimum value, ${\alpha_h} = 1$, in this case.}
 \label{fig:examples}
\end{figure}

There are many possible definitions of a hull for a system defined on a lattice. 
We choose to use the one in which the hull perimeter is 
the number of sites that are on the external perimeter of a domain,
see Fig.~\ref{fig:sketch}, matching the nomenclature in~\cite{Stauffer1994}.

\subsection{Scaled scatter plots}
\label{subsec:scatter}

We analyze scatter plots of areas vs. perimeters and 
 their scaling properties. The results found in~\cite{Sicilia2007} for the Ising case are extended 
in a natural way to the $q=3,\ 4$ cases as discussed below. Further subtleties are
found in models with $q>4$ for which none of the initial states ($T_0=T_c$ and 
$T_0 \to\infty$) are critical. 

\begin{figure} 
 \includegraphics[width=8.5cm]{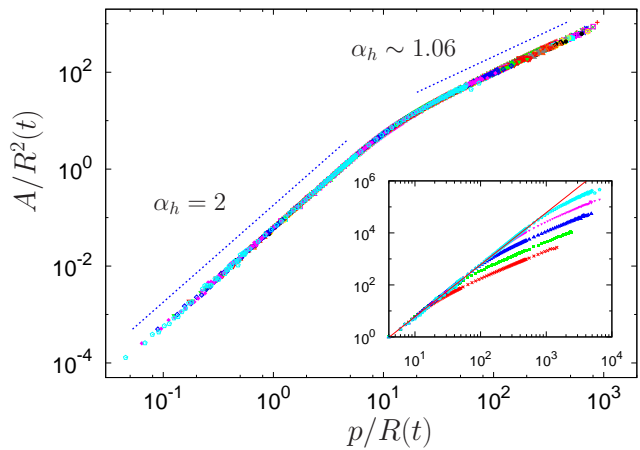}
 \caption{(Color online.) Scaling relation between  hull-enclosed areas and hull 
 perimeters at 
several measuring times for $q=3$ after a quench from $T_0 \rightarrow \infty$ to $T_f=T_c/2$. 
The inset shows the raw data at several times from $t=0$ to $2^{14}$ MCs (bottom to top). The extent up to 
which the 
law $A\sim p^2$ (red solid line) applies increases with time, demonstrating that  
the domains become more regular in the course of evolution.} 
 \label{fig:q3th0pa_hpc}
\end{figure}

As in the Ising case, and independently of the initial condition having long or short range 
correlations, we can distinguish two types of 
domains depending on their relation to the characteristic length $R(t)$.  Hull-enclosed structures
 with area $A<R^2(t)$ have a regular form and the area-perimeter relation is $A \sim p^2$ as in 
Eq.~(\ref{eq:Ap2}). Instead, domains 
with area $A>R^2(t)$ exhibit a rough surface and the area-perimeter relation keeps the power 
law form, $A \sim p^{{\alpha_h}}$, although with an exponent ${\alpha_h} < 2$ that depends on 
the initial condition and $q$.
Figure~\ref{fig:examples} shows 
examples of domains -- taken from our simulations -- with different exponents, 
${\alpha_h}=2$~(a), $1<\alpha_h<2$~(b) and $\alpha_h \simeq 1$~(c). Once hull-enclosed areas 
and perimeters are measured 
with respect to the crossover value, itself written in terms of the characteristic length $R(t)$, 
we can write the scaling relation
\begin{equation}
 \label{eq:scale}
 \frac{A}{R^2} \sim \left( \frac{p}{R} \right)^{{\alpha_h}}
 \; , 
\end{equation}
where 
\begin{equation}
 {\alpha_h} \left\lbrace \begin{array}{lll}
                        = 2 & \;\;\;\; \mbox{for} & A<R^2(t) \; , \\
                        < 2 & \;\;\;\; \mbox{for} & A>R^2(t) \; .
                       \end{array}
          \right.
\end{equation}
Figure~\ref{fig:q3th0pa_hpc} shows the rescaled relation (\ref{eq:scale}) for $q=3$ after a 
quench from $T_0 \rightarrow \infty$ to $T_f=T_c/2$ and several measuring times 
($t=0,2,\ldots,2^{14}$ MCs). We used the characteristic length $R(t) \simeq (\lambda_q \ t)^{1/2}$ 
obtained from the analysis of the equal time correlation $C(r,t)$, as in~\cite{Loureiro2010}. 
The data collapse confirms that $A$ scales as $R^2$ and $p$ as $R$.
We shall see below that this applies to all $q$ and to quenches from different initial conditions.
While for small areas, $A<R^2(t)$, the exponent can be 
obtained from a fit performed on a conveniently long interval, for $A>R^2(t)$ the 
fitting interval is rather
short. Nevertheless, the exponent obtained, ${\alpha_h} \simeq 1.06$, is so far from the 
regular value $2$, that we can definitely conclude that a change in morphology operates upon 
the domains at the crossover determined by $R(t)$. The domains with small areas are regular 
and $A\simeq p^2$. The ones with large areas are rough and 
stretched, the hull perimeters being very close to the hull-enclosed 
area itself, $A\simeq p$. The inset shows the raw data for 
the area $A$ versus the hull perimeter $p$ for $t=0$ to $t=2^{14}$ MCs (bottom to top). 
This figure confirms that 
during evolution the domains grow and become more regular. The red solid line is a 
guide-to-the-eye and represents $A\sim p^2$. 

\begin{figure} 
 \includegraphics[width=8.5cm]{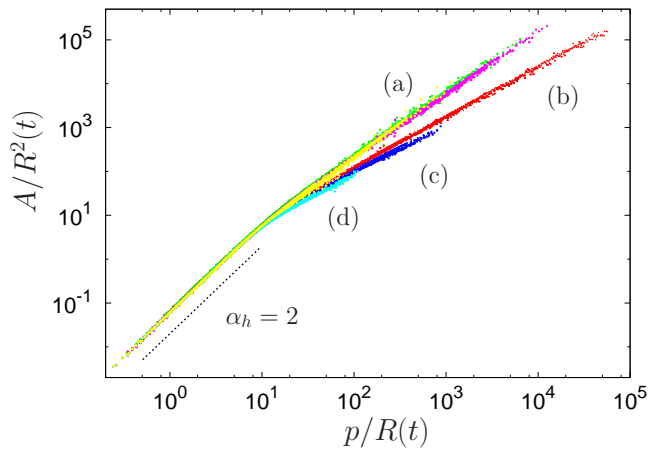}
 \caption{(Color online.) Scaling relation between  
 hull-enclosed areas and hull perimeters at 
$t=2,\ldots,2^{10}$ MCs for different models (various $q$ values) and initial conditions. The curves labeled 
(a) correspond to $q=2, 3, 8$ and $T_0=T_c$, those labeled (b) are for 
$q=2$ and $T_0 \rightarrow \infty$, the ones labeled (c) have been obtained using 
$q=3$ and $T_0 \rightarrow \infty$ and, finally, the ones carrying the label (d) are 
for  $q=8$ and $T_0\rightarrow \infty$. All quenches are to $T_f=T_c/2$. 
When the initial state is at the critical point, 
case (a), the relation $A\simeq p^{{\alpha_h}}$ presents a similar exponent for all $q$, 
${\alpha_h} \simeq 1.44$. Instead, after a quench from the $T_0\rightarrow\infty$ 
uncorrelated state the exponent ${\alpha_h}$ decreases from its $q=2$ value 
${\alpha_h}\simeq 1.15$ to ${\alpha_h} \simeq 1.06$.  Notice that for the
case $q=8$ and $T_0=T_c$, case (a), the observed exponent is 1.44 because the measure was
taken at a relatively small time, in which the finite correlations present at the initial 
state are still large compared with the characteristic lenght. At later times, this exponent will 
decrease, eventually approaching 1.06, as the characteristic lenght grows beyond those initial
correlations.
}
 \label{fig:allpa_hpc}
\end{figure}

Next, in Fig.~\ref{fig:allpa_hpc} we compare  
the relation between hull-enclosed areas and hull perimeters for
systems with different values of $q$ quenched from different initial conditions. 
In all cases we find a power law relation, $A\simeq p^{\alpha_h}$, with varying values of 
${\alpha_h}$. The group of curves labeled (a) corresponds to $q=2, \ 3$ and 8 quenched
from $T_0=T_c$ to $T_f=T_c/2$ and the fit yields ${\alpha_h} \sim 1.44$. 
Interestingly, even the case $q=8$ with short-range correlation
presents an exponent $\alpha_h > 1$. On the other hand, when the initial 
state is one of infinite temperature, the exponent  ${\alpha_h}$
is smaller and decreases with increasing number of states. 
In the first case (b) for $q=2$ we find  ${\alpha_h}\sim 1.12$. For larger values of $q$ it is harder
to be conclusive about the actual value of the exponent since the number of domains with area 
larger than $R^2$ decreases with increasing $q$ and the fitting interval gets smaller. Within our 
numerical accuracy, for $q=3$~(c) ${\alpha_h} \sim 1.06$ and for $q=8$~(d) 
${\alpha_h} \lesssim 1.06$ but this value is  not precise enough to be conclusive. 

Summarizing, in all cases  the characteristic length $R$ grows as $t^{1/2}$, 
the structures become more regular, and the power law $A \sim p^2$ for 
$A<R^2$ is more evident. Concomitantly, 
less domains have an area larger than $R^2$. For $T_0\to\infty$ the value of the exponent 
$\alpha_h$ characterizing the shape of the large hull-enclosed areas 
decreases with increasing $q$, consistent with $\alpha_h\to 1$ for $q\to\infty$. 
For $T_0=T_c$ the exponent $\alpha_h$ takes a higher 
value though still smaller than $2$ for large areas and, within our numerical precision, no 
dependence on $q$ for $q=2,\ 3$ and 8 was observed.


\section{The von Neumann-Mullins law}
\label{sec:rateofchange}



The von Neumann-Mullins law, Eq.~(\ref{eq:dAdt}), predicts that each hull-enclosed area has a 
different rate of change 
depending only on its number of sides (and not on its area). This equation is not fully 
general; it has been derived under certain assumptions that include, as in the Allen-Cahn
case~\cite{AllenCahn}, the fact that the domain wall should be close to flat away from the triple 
points for $q>2$, a feature that can 
be achieved at long times and for long interfaces only. The law ruling the dynamics of small 
areas with highly curved interfaces or large areas with very rough walls is not known 
in general (in the Ising case, however, some exact results for small areas are 
known~\cite{Domany1990,Chayes1995}). 
In this Section we numerically analyze the validity of the von Neumann-Mullins law 
by putting Eq.~(\ref{eq:dAdt}) and its average over small ($A_0<A< cR^2$) or large ($A>cR^2$)
areas to the  test ($A_0$ is a small area cut-off and $c$ a tunable parameter).
We estimate the individual area time-derivate as $A_i(t+1)-A_i(t)$.

\begin{figure} 
\includegraphics[width=8cm]{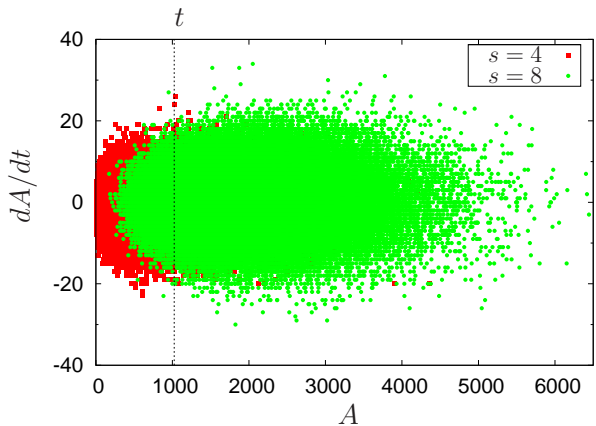} 
 \caption{(Color online.) Rate of change of the individual hull-enclosed area
 at time $t=1024$ MCs after a quench from $T_0\to\infty$ to $T_c/2$ ($q=8$). The data are 
 grouped by the number of sides $n=4$ (red, or dark grey) and $n=8$ (green, or light grey). 
The vertical dashed line indicates $R^2(t) \propto t$.}
 \label{fig:cloud}
\end{figure}

In Fig.~\ref{fig:cloud} we examine Eq.~(\ref{eq:dAdt}) by plotting $dA/dt$ vs $A$ for $n=4$ 
and $n=8$ both at $t=1024$ MCs in a $q=8$ Potts model 
quenched from $T_0\to\infty$ to $T_c/2$ (similar  ellipsoidal  
clouds are obtained for $T_0=T_c$ and for $q=3$).
The vertical dashed line
shows $R^2(t) \propto t$. The large vertical spreading of the data points let us conclude that the 
von Neumann-Mullins law does not hold strictly in our case, 
contrary to what was found 
in a phase field simulation of ideal grain growth~\cite{KiKiKiPa06} and in a Potts models with very 
large $q$ and microscopic updates modified to ensure a local $q=6$ constraint and thus
accelerate the evolution~\cite{Kim10}. 

\begin{figure} 
 \includegraphics[width=8cm]{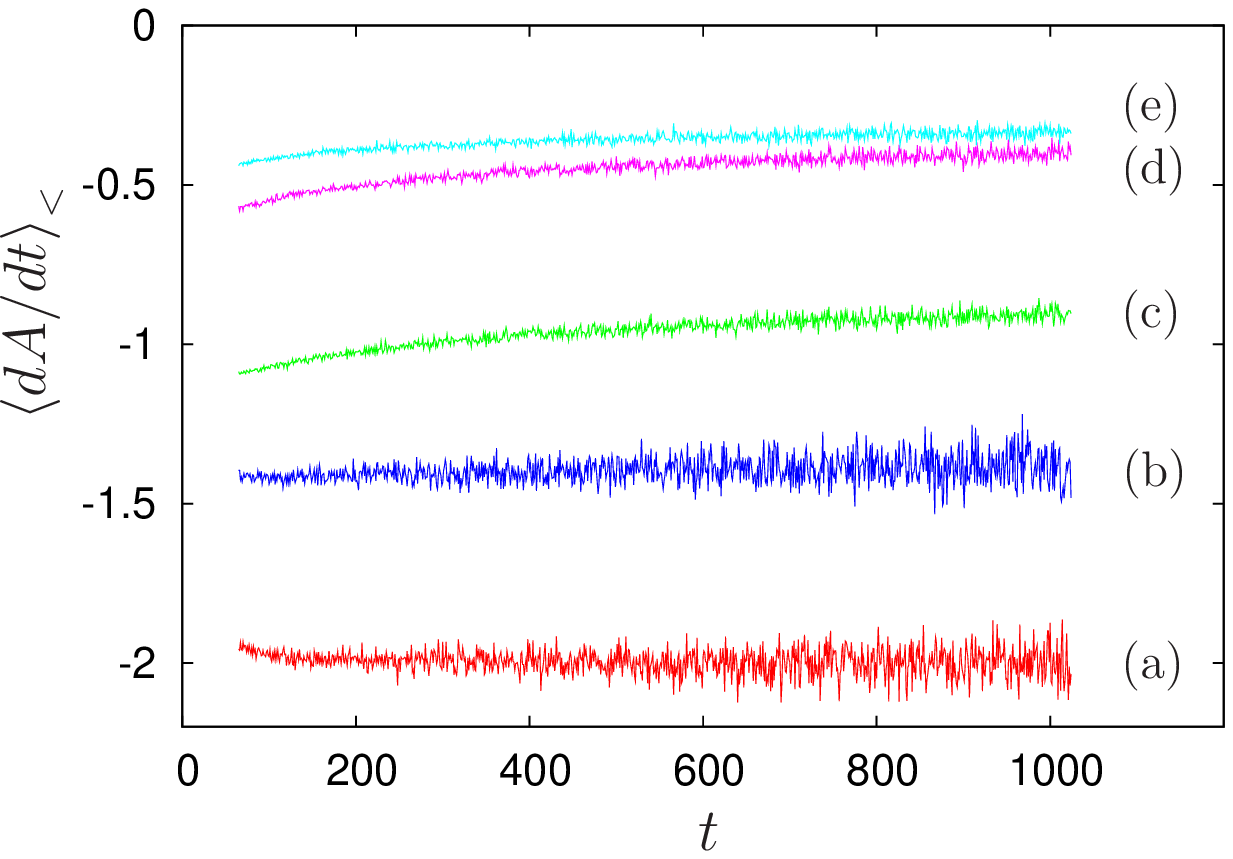}
 \includegraphics[width=8cm]{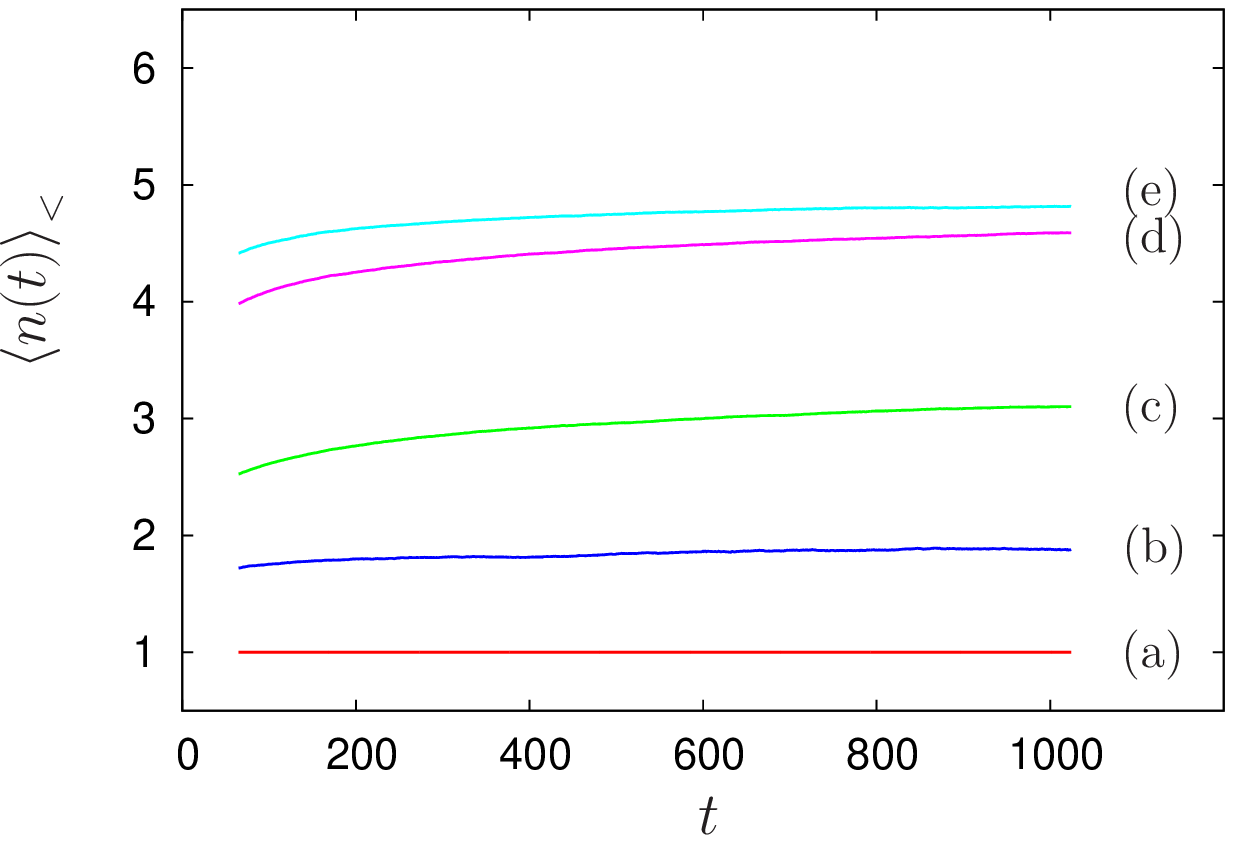} 
 \caption{(Color online.) Top panel: The area rate of change averaged over small areas,
 $A_0=8<A<cR^2$, for 
(a) $q=2$ and $T_0\to\infty$ (red),
(b) $q=3$ and $T_0=T_c$ (blue),
(c) $q=3$ and $T_0\to\infty$ (green), 
(d) $q=8$ and $T_0=T_c$ (pink), and  
(e) $q=8$ and $T_0\to\infty$ (cyan). 
In all cases the averaged rate of change seems to asymptotically reach a negative constant. 
Bottom panel:  The time-dependent
average number of sides, $\langle n(t) \rangle_<$
for the same parameters is shown with the same color and label code. 
See the text for the discussion.}
 \label{fig:rcALL_asmall}
\end{figure}

Figure~\ref{fig:rcALL_asmall} (top panel) shows the area rate of change averaged 
over small areas:
\begin{equation}
\left\langle \frac{dA}{dt} \right\rangle_< 
= 
\frac{1}{N_{<}(t)} \sum_{A_0<A_i<cR^2} \frac{dA_i}{dt} 
\end{equation}
where $N_<(t)$ is the number of areas obeying $A_0<A_i<cR^2$.
When one area leaves the defining interval $[A_0,cR^2]$, 
it is no longer taken into account in the sums; 
on the other hand, new areas can enter this interval.
These are the reasons for the time-dependence in $N_<$. 
Each curve in the figure
corresponds to different values of $q$ and temperature of the initial conditions: 
(a) $q=2$ at $T_0\to\infty$, 
(b) $q=3$ at $T_0=T_c$, 
(c) $q=3$ at $T_0\to\infty$, 
(d) $q=8$ at $T_0=T_c$ and 
(e) $q=8$ at $T_0\to\infty$. 
In cases (a) and (b) the averaged area rate of change has reached a constant 
(within the fluctuations). In the  other cases, although the average continues to 
evolve within the explored time window, 
the variation is much smaller than the total mean area. 
In all cases $\langle dA/dt \rangle_<$ is negative and, according to the 
von Neumann-Mullins's law, this implies an average number of sides  smaller than 6.
 The fluctuations observed mainly in 
the case $q=3$ are the result of coalescence and dissociation processes that become increasingly 
rare as the number of ground states of the system increases. 

\begin{figure} 
 \includegraphics[width=8cm]{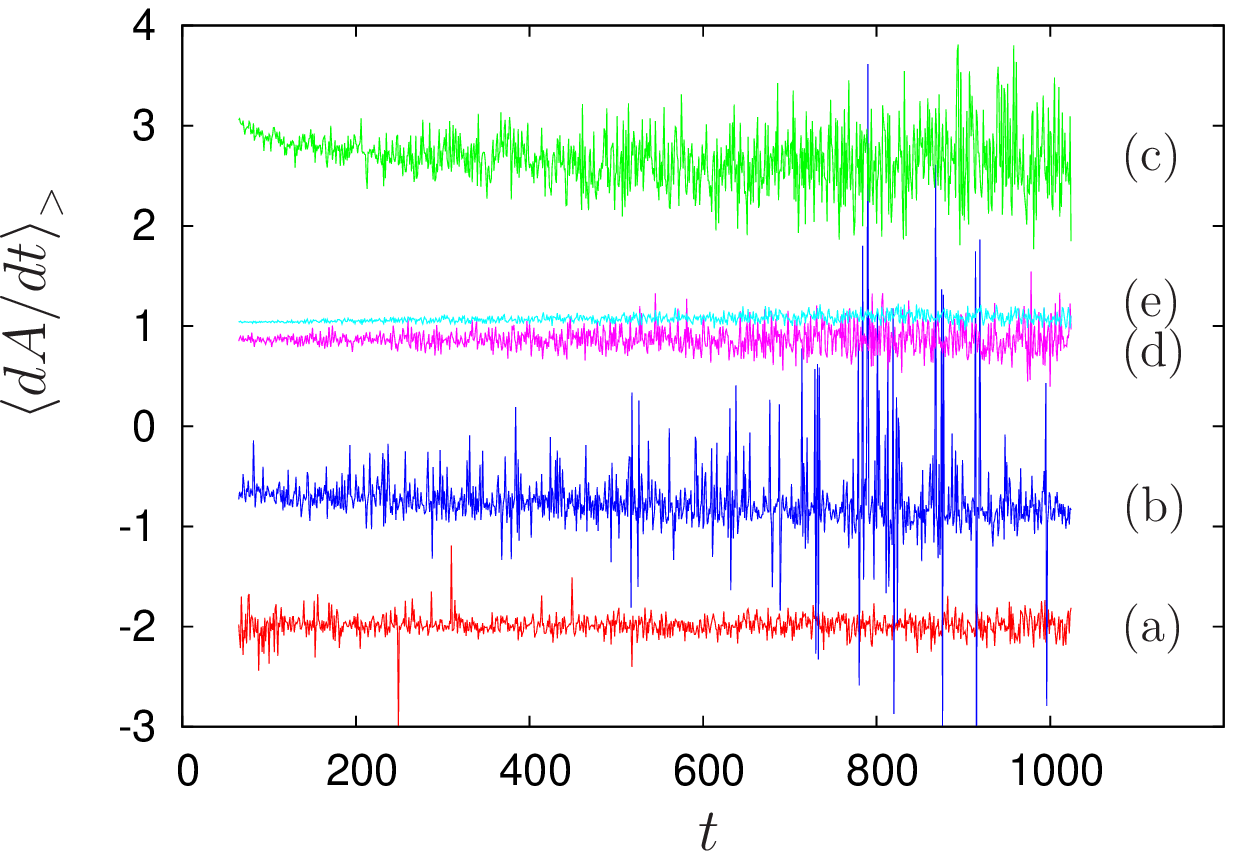}
 \includegraphics[width=8cm]{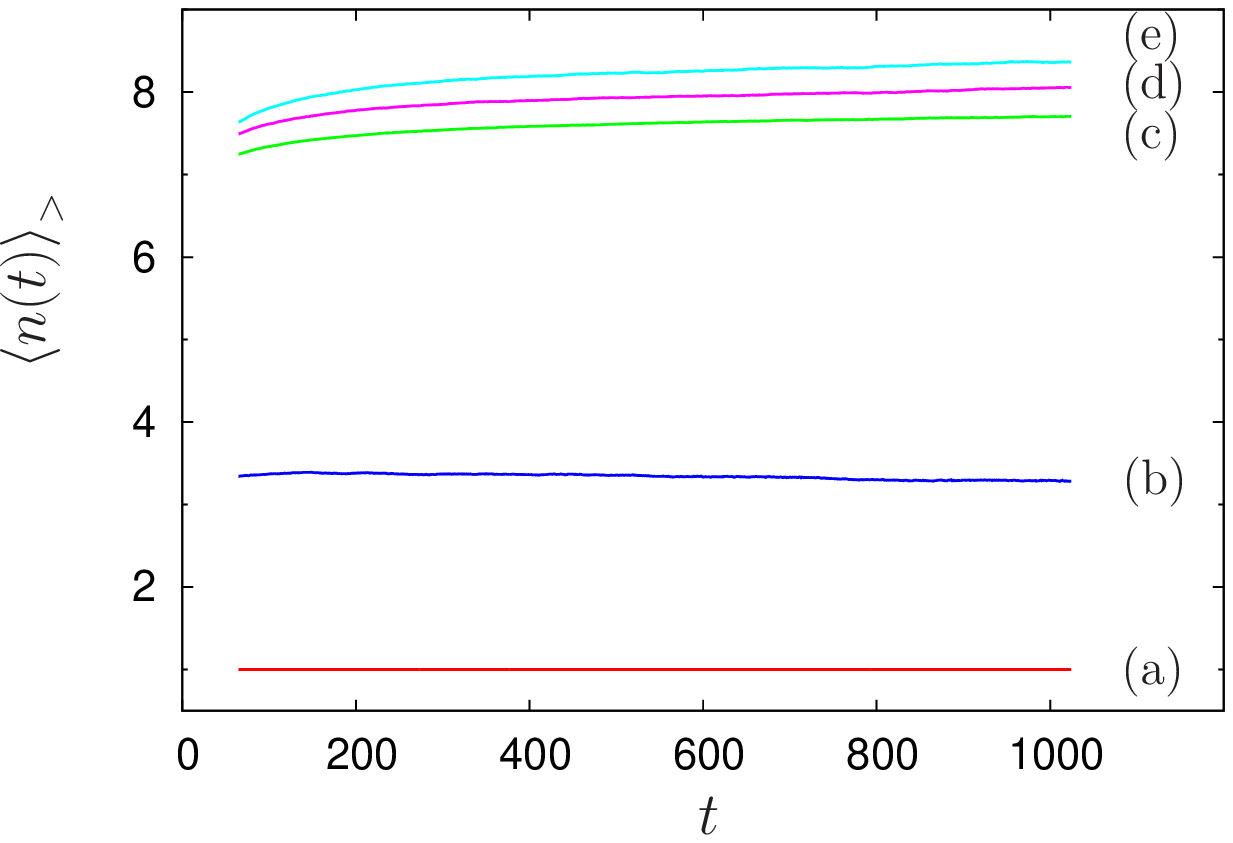} 
 \caption{(Color online.) The same as in Fig.~\ref{fig:rcALL_asmall} but
 averaging over large areas, $A>cR^2$, for the same choice of parameters
 and color code used in that figure. Notice the order of the labels in
the top panel.}
 \label{fig:rcALL_alarge}
\end{figure}

The results in the top panel of Fig.~\ref{fig:rcALL_asmall} are to be compared 
to those in the bottom panel, where
the small-area average of the time-dependent
number of sides, $\langle n(t) \rangle_<$, is displayed. 
The results are qualitatively 
and quantitatively consistent with the averaged von Neumann-Mullins law 
\begin{equation}
\left\langle \frac{dA}{dt} \right\rangle_< 
=
\frac{\lambda_q}{5} \left( \langle n(t) \rangle_< - 6 \right) 
\; .
\end{equation}
The values for $\lambda_q$ found by comparing the two sets of curves are
consistent with $\lambda_q$ decreasing for increasing $q$ and independent of the
initial temperature $T_0$, as is also found from the analysis of the space-time 
correlation function~\cite{Loureiro2010}. 

The same analysis performed on the large areas is much more delicate. To 
start with, it is difficult to collect good statistics since only a
few sufficiently large areas remain in the samples at long times, 
especially for large values of $q$. Figure~\ref{fig:rcALL_alarge} 
present data for the same set of parameters as in Fig.~\ref{fig:rcALL_asmall}.
Clearly, the fluctuations in the averaged area rate 
of change are much larger than when the average was performed over small 
areas.  Still, one can argue that the rate of change asymptotically approaches a constant 
  in all cases. However, and contrary to the small 
area case, the constant is negative only in the cases in which 
the initial state is critical ($q=3$ and 4, not shown, at $T_0=T_c$) 
and for the special case $q=2$ with $T_0\to\infty$. In all other cases 
[$q=3$, $T_0\to\infty$ (c), $q=8$, $T_0=T_c$ (d) and $q=8$, $T_0\to\infty$ (e)] 
the constant is positive. In cases (a) and (b) one finds agreement between the 
negative averaged area rate of change and the fact that the averaged number of sides
of these areas is smaller than 6. Moreover, the $\lambda_q$ values
found, within our numerical accuracy, are consistent with
the ones stemming from the analysis of the data averaged
over small areas. 

The remaining three cases have to be discussed
separately. While the signs are consistent, i.e. positive averaged area rate of change and 
average number of sides larger than 6, the trend of these curves is not systematic and 
shows that such large structures do not follow the linear behavior predicted by the
von Neumann-Mullins law.
This might be due to different reasons. For instance, in the case $q=3$ with $T_0\to\infty$ (c) 
we ascribe this feature to the fact that the long-time domain structure is special in this case, with 
many large domains with a rather rough surface, as shown in panel (e) of Fig.~\ref{fig:snapshotq3} 
and confirmed by the fact that $\alpha_h$ is very close to one, see Fig.~\ref{fig:q3th0pa_hpc}, in 
this regime. 

\begin{figure} 
 \includegraphics[width=8cm]{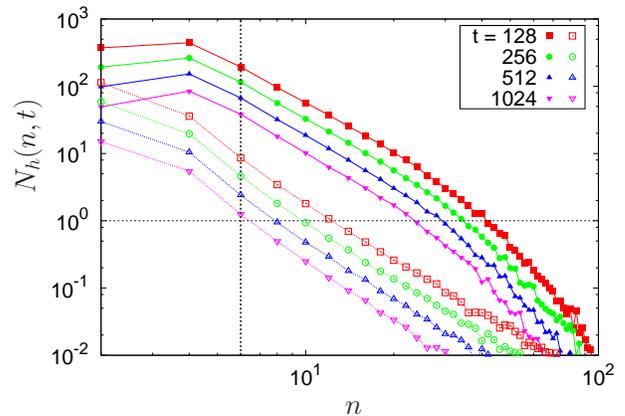}
 \caption{(Color online.) Total number of hulls with $n$ sides for $q=3$,  $T_0\to\infty$
(solid symbols, upper group of curves) and $T_0=T_c$ (empty symbols, lower group of curves). 
The dotted lines indicate $N_h=1$ (horizontal) and $n=6$ (vertical). The different symbols 
correspond to different measuring times as given in the key. 
The distribution has weight well above $N_h=1$ for $n>6$ when the initial conditions have finite 
correlation length.}
 \label{fig:nsides_allq3log}
\end{figure}

Figure~\ref{fig:nsides_allq3log}  display 
the total number of hulls with  $n$ sides, $N_h(n,t)$, in a
log-log scale, showing the behavior at four different 
times, for $q=3$ after a quench from $T_0 \rightarrow \infty$ (solid 
symbols) and $T_0=T_c$ (empty symbols) to $T_f=T_c/2$. The dotted horizontal and vertical lines 
indicate $N_h=1$ and $n=6$, respectively. For  initial conditions with finite 
correlation length (upper group of curves)  many interfaces have more than 6 sides and 
the average value $\langle n\rangle$ is less representative of the fluctuating behavior. The 
distribution tail seems exponential. Instead, for critical initial conditions (lower group of curves), the number of hulls with 6 or more sides becomes 
negligible during evolution  and the distribution decays as a 
power law.  
Interestingly, for $T_0\to\infty$ the 
curves are monotonically decreasing for $n\geq 4$ while for $T_0=T_c$ they are 
monotonically decreasing for $n\geq 2$ (note that for the case 
$q=3$ shown here, $n$ only takes even values). 
The former feature is experimentally established although it is not captured by simplified random 
neighbor models~\cite{Flyvbjerg93,Durand10}.

\begin{figure}  
 \includegraphics[width=8cm]{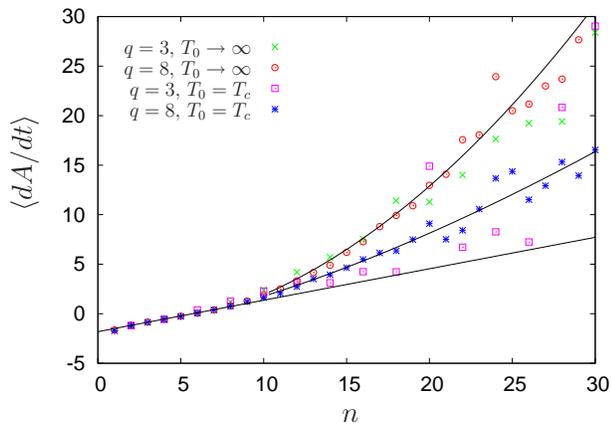}
 \caption{(Color online.) 
 The full averaged area rate of change as a function of the number 
 of sides at $t=64$ MCs for the parameters given in the key. The solid lines correspond to the 
 function $f(x)\simeq x^\gamma$ with $\gamma=1, \ 1.5, \ 2$, from bottom to top, and
serve as guides to the eyes.}
 \label{fig:averagedA-vs-n}
\end{figure}

The summary of our analysis of the von Neumann-Mullins law is given in 
Fig.~\ref{fig:averagedA-vs-n} where we averaged 
the area rate of change over all areas with given value of $n$ and displayed 
the results as a function of $n$ for the $q=3$ and $q=8$ models quenched from 
$T_0=T_c$ and $T_0\to\infty$.  The data are measured at $t=64$ MCs.
Although the $n-6$ dependence is observed for both values of $q$ and the
different initial conditions, 
the behavior is linear in the averaged data only up to $n\simeq 10$ (lower solid straight 
line). Deviations appear for larger values of $n$ in all cases.

\section{Hull perimeter distributions}
\label{sec:distribution}

In a previous work, Ref.~\cite{Loureiro2010}, the hull-enclosed area distribution, $n_h(A,t)$, was studied for the $q$-state Potts model after
temperature quenching the system from above the critical temperature into the ordered phase. We now describe the
time evolution of the hull perimeter distribution, $n_h(p,t)$, following this quench and relate it to $n_h(A,t)$. The form
of the distribution depends  on the initial condition and on the value of $q$. In particular, there is a dependence
on the morphological characteristics of each domain, that are different for structures with large or small sizes, when
compared to the characteristic lenght $R(t)$. 

\subsection{Critical initial condition}


\begin{figure}  
 \includegraphics[width=8cm]{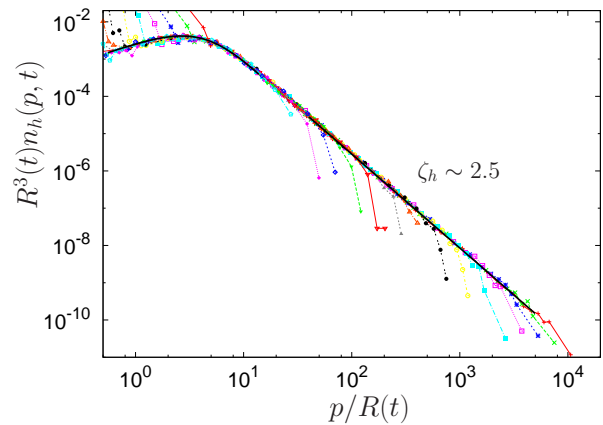}
 \caption{(Color online.) 
 Rescaled hull perimeter distribution for $q=3$ at several times ($t=2,2^2,\ldots,2^{14}$~MCs),
 after a quench from $T_0=T_c$ to $T_f=T_c/2$. The solid black line represents the theoretical
 prediction, Eq.~(\ref{eq:perimdist}), valid for small areas, $A/R^2 < 10$ with $\alpha_h = 2$
 and for large areas, $A/R^2>10$, with $\alpha_h \simeq 1.44$.} 
 \label{fig:q3th500phl_hpc}
\end{figure}

For $2\leq q\leq 4$ the transition is continuous and the initial state at $T_0=T_c$ is critical; hence
all initial distributions, i.e. of areas and perimeters, should have a power law tail. 
The functional form of the equilibrium number density of hull-enclosed areas at $t=0$ is $n_h(A,0) = (q-1) c^{(q)}_h/A^2$~\cite{Loureiro2010}. 
Since one and only one hull is associated to 
each hull-enclosed area,  the number density of hull perimeters should be linked to the one of 
hull-enclosed areas according to $n_h(p,0)dp=n_h(A,0)dA$. Using $A\simeq p^{\alpha_h}$, a
relation that is valid one-to-one, one finds 
\begin{equation}
 n_h(p,0) \sim \frac{\gamma_q}{p^{\zeta_h}}
\label{eq:dist-p}
\; .
\end{equation}
where $\gamma_q = \alpha_h (q-1) c^{(q)}_h$ and $\zeta_h = \alpha_h +1$. 
The exponent $\zeta_h$ is the equivalent to the Fisher exponent $\tau$ for the area distribution 
now describing the perimeter distribution. 
In two dimensions $\zeta_h$  is linked to the fractal dimension of the perimeter
$D_h$ as \cite{Stauffer1994}
$ \zeta_h = 1+2/D_h$.
In the special case $q=2$ at $T_c$ one has $D_h=11/8$~\cite{Vanderzande1989}. 

Using the fact that areas and perimeters are related one to one also dynamically,
$n_h(p,t) dp=n_h(A,t)dA$, and the power-law relation between areas and perimeters, Eq.~(\ref{eq:scale}),
the time-dependent number density of hull perimeters can be easily written as 
\begin{equation}
 n_h(p,t) \sim \dfrac{\gamma_q}{R^3} \; 
 \dfrac{\left(\dfrac{p}{R}\right)^{\alpha_h-1}}{\left[1+\left(\dfrac{p}{R}\right)^{\alpha_h} \right]^2 }
 \; .
 \label{eq:perimdist}
\end{equation}
The value of $\alpha_h$ depends on which of the   
two regimes, $p<R$ and $p>R$, one focuses on, with $\alpha_h =2$ in the former and $\alpha_h<2$
in the latter. A similar argument was used in~\cite{Sicilia2007} and \cite{Sicilia2009} to obtain the perimeter
number density in the $2d$ Ising model with non-conserved and conserved order parameter dynamics,
respectively.

Figure~\ref{fig:q3th500phl_hpc} shows the hull perimeter number density, scaled by the 
characteristic length $R(t)$, for the critical model with $q=3$ after a quench 
to $T_f=T_c/2$.  
The long-range correlations present in the critical initial state 
are preserved during evolution and  lead to the power law tail in the distribution.
The exponent of this tail is $\zeta_h \sim 2.5$ 
in agreement with the theoretical estimate
$\zeta_h = \alpha_h +1 \simeq 1.44 + 1$, see Fig.~\ref{fig:allpa_hpc}. For small areas, $p\ll R$, 
one finds $R^3 n_h(p,t) \simeq \gamma_q p/R$ also in good agreement with the 
analytic prediction. Thus, equation~(\ref{eq:perimdist}) describes well these two limits.

\subsection{Non-critical initial condition}
\label{section.noncritical}

\begin{figure}
 \includegraphics[width=8cm]{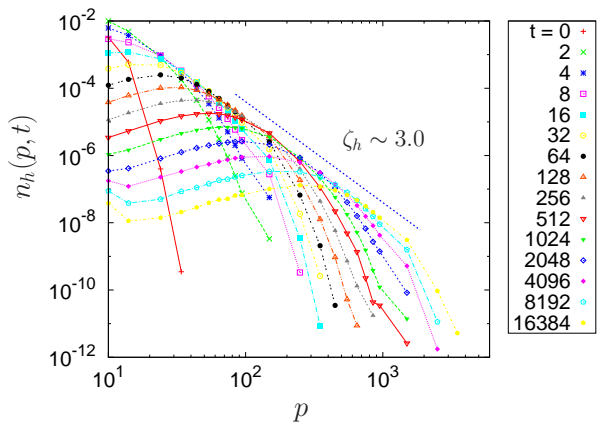}
 \caption{(Color online.) Hull perimeter distribution for $q=8$ at several times given in the key  
 after a quench from equilibrium at $T_0 \rightarrow \infty$ to $T_f=T_c/2$. The declivity of the 
 envelope, $\zeta_h \sim 3.0$, is a direct consequence of dynamical scaling.} 
 \label{fig:q8th0phl_hp}
\end{figure}

Being far from critical percolation, the initial state at $T_0\to\infty$ for a system with $q>2$
does not become critical during its evolution and the area (and consequently the perimeter)  distribution does not develop a power law tail.
In Fig.~\ref{fig:q8th0phl_hp} we use log-log scale to show the $q=8$ hull perimeter 
distribution after a quench from $T_0 \rightarrow \infty$ to $T_f=T_c/2$ at several times.  
Similarly to what happens with the hull enclosed area distribution for 
$q>2$~\cite{Loureiro2010}, there is a $p^{-3}$ envelope that is a consequence of dynamical 
scaling. 

\begin{figure} 
\includegraphics[width=8cm]{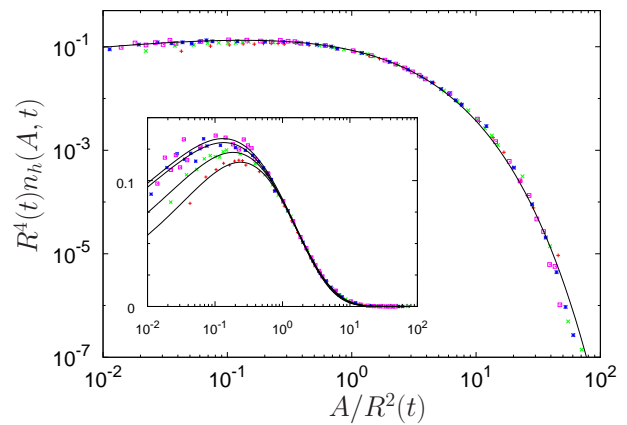}
\caption{(Color online.) Hull-enclosed area distribution for $q=8$ at several times  
after a quench from equilibrium at $T_0 \rightarrow \infty$ to $T_f=T_c/2$,
rescaled by $R^2(t)$.  The solid line is a fit with Eq.~(\ref{eq.gamma})
and $a\simeq 1.21$, $b\simeq 1.22$, $c\simeq 0.6$. Inset: the same data in
a linear scale, showing that for small areas, the collapse improves at large
times.}
\label{fig:gamma_q8_area}
\end{figure}

For very large $q$, after the quench, domains start to increase from localized
density fluctuations that are randomly scattered throughout the system, in a similar way to
the Avrami-Johnson-Mehl method~\cite{FeNe07} used to produce Voronoi diagrams. Thus, neglecting 
coalescence effects that are only important for small $q$, it can be conjectured
that the set of domain centers (defined, for example, as the center of mass of a domain) 
may be approximately described as a set of random points from which Voronoi domains are
drawed.
Indeed, as we show in Fig.~\ref{fig:gamma_q8_area}
for $q=8$ (see also Ref.~\cite{Loureiro2010}), the distribution of areas after
a temperature quench from $T_0\to\infty$ to $T_c/2$, is 
well described by the generalized Gamma (or Voronoi) distribution~\cite{FeNe07}, widely
used in grain growth literature~\cite{Norbakke02,Wang03,Xiaoyan00,Kim10}:
\begin{equation}
n_h(A,t) \sim 
\left( \frac{A}{R^2}\right)^{a-1}
\exp\left[-b\left( \frac{A}{R^2}\right)^c \right],
\label{eq.gamma}
\end{equation}
with $a\simeq 1.21$, $b\simeq 1.22$, $c\simeq 0.6$. Simpler versions of the above function
have also been proposed, with two ($c=1$) or a single ($c=1$ and $a=b$) parameter. Indeed,
the values that we get for $a$ and $b$ are almost the same, but $c\neq 1$.

Once the distribution of areas is known, in principle one could relate it, as in
the preceeding section, to the perimeter distribution using Eq.~(\ref{eq:scale}), obtaining,
once again, a Gamma distribution:
\begin{equation}
n_h(p,t) \sim R \left( \frac{p}{R}\right)^{\alpha_h a -1} \exp\left[ -b\left( \frac{p}{R} \right)^{\alpha_h c} \right].
\label{eq:np}
\end{equation}
Differently from the area distribution, the dependence on $\alpha_h$ characterizes two distinct regimes, $\alpha_h=2$ 
for $p<R$ and $\alpha_h <2$ for $p>R$. As one can see from Fig.~\ref{fig:q8th0phl_hp}, the data are well
described by the above distribution. However, the fit parameters are not only different from those
obtained from the area distribution but differ as well between the regions with small and
large $p$. 
One possible reason is that although 
Eq.~(\ref{eq.gamma}) is a good approximation, the dynamics of a growing domain is not equivalent to 
the Avrami-Johnson-Mehl method. If the Voronoi cells of the domain centers are drawn, 
there will be a large superposition with the actual domains, but those regions close to the interfaces
could be wrongly assigned because, differently from Voronoi diagrams in which the borders are as
straight as they can be on a lattice, a coarsened domain has rougher interfaces.
Thus, the deviations are on the scale of the perimeter lenght and, although they may not be noticeable for 
area distributions, they become visible in the perimeter distributions.

\begin{figure} 
\includegraphics[width=8cm]{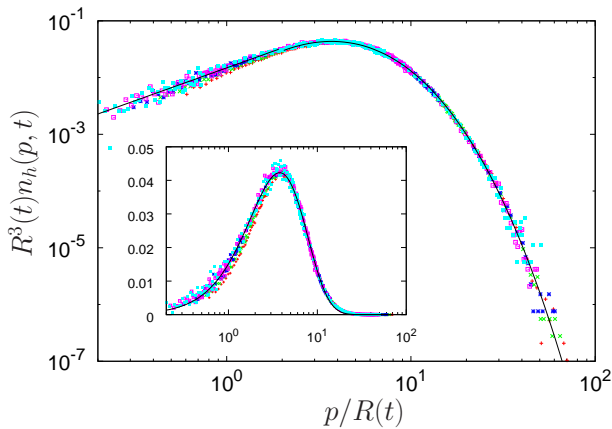}
\caption{(Color online.) Hull perimeter distribution for $q=8$ at several times  
after a quench from equilibrium at $T_0 \rightarrow \infty$ to $T_f=T_c/2$,
rescaled by $R(t)$. The solid line is Eq.~(\ref{eq:np}) with $\alpha_h = 2$
for $p<R$ and $\alpha_h\simeq 1.06$ for $p>R$. The fit parameters, however,
besides being different from those of Fig.~\ref{fig:gamma_q8_area}, differ
also for $p<R$ and $p>R$. Inset:  the same data in a linear scale.}
\label{fig:gamma_q8}
\end{figure}


\section{Conclusions}
 \label{sec:conclusions}

We studied several geometric properties of the nearest-neighbour
$q$-state Potts model quenched below its phase transition. 
In particular we analyzed the von Neumann-Mullins law, 
the hull perimeter length distribution and 
the relation between hull perimeters and their enclosed areas. 
Diversely from the so called cellular Potts model, in which several 
layers of interacting neighbours are considered in order to decrease
the lattice anisotropy, here it is an important ingredient. Moreover,
coalescence is also relevant when $q$ is not too large, as the
values considered here. These two factors, necessary for the
single domain von Neumann-Mullins law, are here violated and deviate the model 
from the ideal grain growth conditions.

After having confirmed that all dynamic observables satisfy dynamic scaling with respect to 
the growing length $R(t) \simeq (\lambda_q t)^{1/2}$, we studied the fractal properties 
of the hulls by plotting $A/R^2(t)$ against $(p/R(t))^{\alpha_h}$, Eq.~\ref{eq:scale}. We found that for 
$p/R(t)\ll 1$ interfaces are smooth with $\alpha_h=2$ while for $p/R(t) \gg 1$ they are 
fractal with $\alpha_h$ being smaller for $T_0\to\infty$ than for $T_0=T_c$ and decreasing 
with $q$. Thus, the 
behavior of objects with linear sizes that are smaller  that the typical growing length have a rather 
different dynamics, morphology and statistical properties than those whose linear sizes are 
larger than the dynamical length. 

The linear proportionality between $dA/dt$ and $n$ (independently of $A$),
the von Neumann-Mullins law, does not hold for each individual area for the
nearest neighbour Potts model in the temperature interval considered here
after the quench. This may be due to thermal fluctuations occuring along 
the perimeter that may mask the curvature driven contribution. We then 
examined the law on the mean, by averaging over areas that are smaller or larger than the 
typical one, $R^2(t)$. In the small area regime the linear von Neumann-Mullins law is 
verified on the mean while for larger areas it is not, with deviations from linearity
depending on the value of $q$ and initial conditions. The same separation of regimes is found when averaging
over all areas that have the same number of sides. The behavior at large values of $n$ 
seems to be captured by a modified power law involving $\alpha_h$ but our data are
not extensive enough to test this conjecture. 

Finally, we studied the hull perimeter distribution distinguishing critical ($T_0=T_c$ for 
$q\geq 4$) from non-critical (all other cases, especially large $q$ for $T_0\to\infty$). In the 
former case we found that the scaling function can be derived with an argument that 
combines the use of Eq.~(\ref{eq:scale}) with the form of the 
distribution of hull-enclosed areas found in Ref.~\cite{Loureiro2010}. For non-critical 
initial conditions we found, instead, that the Gamma function commonly used in studies 
of ideal grain growth describes well both the hull-enclosed area and perimeter distributions,
although the expected relation between the exponents of both distributions is not obeyed.
The reason is due to the fact that the perimeter distribution is much more sensitive to
the differences between the actual domains and their Voronoi approximations, as discussed
in section~\ref{section.noncritical}. 
 
There are, however, several statistical properties of interface sides and 
topological properties for cellular systems that were not yet fully studied for the
nearest-neighbour Potts model, and will be the subject of a future publication.


\acknowledgments
We thank M. Picco and T. Blanchard for useful discussions. Work partially supported by
CAPES/Cofecub, grant 667/10. JJA is also partially supported by the 
Brazilian agencies CNPq and Fapergs.


\bibliographystyle{apsrev4-1}
\bibliography{library.bib}

\end{document}